\documentclass[preprint,12pt]{elsarticle}
\newcommand{\R}{{\mathbb{R}}}
\newcommand{\Z}{{\mathbb{Z}}}
\usepackage{amssymb}

\journal{Frontiers in Physics}

\begin{document}
%\sloppy
%\draft
\begin{frontmatter}
\title{ Point interactions with bias potentials
}
\author{A.V. Zolotaryuk$^1$, G.P. Tsironis$^{2,3}$, Y. Zolotaryuk$^1$}
%\email{azolo@bitp.kiev.ua}
\address
{$^1$Bogolyubov Institute for Theoretical Physics, National Academy of
Sciences of Ukraine, Kyiv 03143, Ukraine \\
$^2$School of Engineering and Applied Sciences, Harvard University, Cambridge, MA 02138, USA\\
$^3$Department of Physics, University of Crete, Heraklion 71003, Greece}

\date{\today}

\begin{abstract}
We develop an  approach on how to define single-point interactions under the application of  external fields. The essential feature relies on an asymptotic method
 based on the one-point approximation of multi-layered heterostructures that are
 subject to bias potentials. In this approach , the zero-thickness limit 
 of the transmission matrices of specific structures is analyzed and shown to
 result in matrices  connecting the two-sided boundary conditions
 of the wave function at the origin. The reflection and transmission 
 amplitudes are computed in terms of these matrix elements as well as
 biased data. Several one-point interaction models of two- and 
 three-terminal  devices 
  are elaborated. The typical transistor in the semiconductor physics is 
  modeled in the ``squeezed  limit'' as a $\delta$- and a $\delta'$-potential
  and referred to as a  ``point'' transistor. The basic property of these
  one-point interaction models is the existence of several extremely sharp peaks
  as an applied voltage  tunes, at which the transmission amplitude is non-zero, 
  while  beyond these resonance values, the heterostructure behaves as a fully
  reflecting wall. The location of these peaks referred to as a ``resonance set''
  is shown to depend on both system parameters and applied voltages.
    An interesting effect of resonant transmission through 
  a $\delta$-like barrier  under the presence of an adjacent well is observed. 
This transmission occurs at a countable set of the well depth values.
  \end{abstract}

\begin{keyword} 
one-dimensional quantum systems, transmission, point interactions,  
 resonant tunneling, controllable potentials, heterostructures

\end{keyword}
%\pacs{03.65.-w, 03.65.Nk, 73.40.Gk}
%\maketitle

\end{frontmatter}

%===========================================Introduction================================
\section{Introduction}

One-dimensional
quantum systems modeled by  Schr\"{o}dinger operators with singular zero-range potentials 
have been discussed widely in both the physical and mathematical literature
(see books~\cite{do,a-h,ak} for details and references). 
 Additionally,  a whole body of literature beginning from the early publications
 \cite{s,k,cnp,adk,cnt,an,ni1,ni2}  (to mention just a few)  has been published,  
where  the  one-dimensional  stationary Schr\"{o}dinger equation 
\begin{equation}
- \, \psi''(x) +V(x) \psi(x) = E\psi(x),
\label{1}
\end{equation}
 with the potential $V(x)$ given in the form of distributions, where 
 $\psi(x)$ is the wave function and 
 $E$ the energy of an electron, was shown to 
exhibit a number of peculiar features with possible applications to quantum physics.
Currently, because of the rapid progress in fabricating nanoscale quantum
devices, of particular importance is the point modeling of different structures like quantum
waveguides \cite{acf,ce}, spectral filters \cite{tc1,tc2}  or infinitesimally thin sheets
 \cite{zz15pla,zz15jpa}.

 In the present paper we follow the traditional approach (see the work \cite{adk} by Albeverio {\it et al} and 
 references therein), according  to which 
there exists a one-to-one correspondence between the full set of self-adjoint extensions of the one-dimensional free Schr\"{o}dinger  operator 
and the two families of boundary conditions: {\it non-separated} 
and {\it separated}. The non-separated extensions describe non-trivial four-parameter 
point interactions subject to the two-sided at $x= \pm 0$ boundary conditions on 
the wave function $\psi(x)$ and its derivative $\psi'(x)$ given by the connection matrix of the form 
%------------------------------------1y--------------------------------------------
\begin{eqnarray}
\left( \begin{array}{cc} \psi(+0)  \\
\psi'(+0) \end{array} \right) 
 = \Lambda \left(
\begin{array}{cc} \psi(-0)   \\
\psi'(-0)   \end{array} \right), ~~~~ \Lambda = {\rm e}^{{\rm i} \chi}
 \left( \begin{array}{cc} \lambda_{11}~~ 
\lambda_{12}\\
\lambda_{21} ~~ \lambda_{22} \end{array} \right) ,
\label{1y}
\end{eqnarray}
%------------------------------------1y------------------------------------------
 where $\chi \in [0, \, \pi ),~\lambda_{ij} \in \R$ fulfilling the condition $\lambda_{11} \lambda_{22}- \lambda_{12}\lambda_{21} =1$. The separated point interactions are described by the direct sum of the free Schr\"{o}dinger operators defined on the half-lines $(-\infty, \, 0)$, $(0,\, \infty)$ and subject to the following pair of boundary conditions: 
%----------------------------------2y-------------------------------------
\begin{equation}
 \psi'(-0)= h^{-} \psi(-0)~~~~\mbox{and}~~~~ \psi'(+0)= h^+ \psi(+0),
\label{2y}
\end{equation} 
%--------------------------------------2y-------------------------------------
where $h^\pm \in \R \cup \{\infty \}$. For instance, if $\{h^-, \, h^+\} = 
\{\infty, \, \infty\}$,  Equations (\ref{2y}) describe the Dirichlet boundary conditions
$\psi(\pm \, 0) =0$. 
In physical terms, a separated self-adjoint extension means that the corresponding point potential 
is completely opaque for an incident particle. Alternatively,  the boundary conditions
can be connected using the Asorey-Ibort-Marmo formalism \cite{aim} or the 
Cheon-F\"{u}l\"{o}p-Tsutsui approach \cite{cft,tfc}. The advantage of both these connecting
representations is that they enable to include all the self-adjoint extensions without 
treating the particular cases as  any parameters tend to infinity. In other words, the relations 
(\ref{2y}) are excluded from the consideration.

Some particular examples of Equation (\ref{1}) and  the corresponding 
$\Lambda$-matrix (\ref{1y}) are important in applications. The most simple 
and widespread  potential is Dirac's delta function $\delta(x)$, i.e.,
$V(x) = \alpha \delta(x)$ where $\alpha$ is a strength constant (or intensity). 
The wave function $\psi(x)$ for this interaction (called the $\delta$-interaction or $\delta$-potential) 
is continuous at the origin $x =  0$, whereas its derivative undergoes a jump, so that the boundary conditions 
read $\psi(-\, 0)= \psi(+\, 0) =: \psi(0)$ and $\psi'(+\,0) - \psi'(-\, 0) =\alpha \psi(0)$ yielding 
the $\Lambda$-matrix in the form 
\begin{equation}
\Lambda = \left(\begin{array}{cc} 1~~~0 \\ \alpha ~~~ 1 \end{array} \right).
\label{1u}
\end{equation}
In the simplest case, this point potential is constructed from constant functions defined on 
a squeezed interval. 

The dual point interaction for which  the derivative $\psi'(x)$ is continuous at the origin, 
but $\psi(x)$
discontinuous,  is called a $\delta'$-interaction (the notation adopted in the literature 
\cite{a-h}). 
This point interaction with strength  $\beta$ defined by the boundary
conditions $\psi'(-\,0) = \psi'(+\, 0) =: \psi'(0)$ and $\psi(+\, 0) = \psi(-\, 0)= 
\beta \psi'(0)$
 has the $\Lambda$-matrix of the form
\begin{equation}
\Lambda = \left(\begin{array}{cc} 1~~~\beta \\ 0 ~~~ 1 \end{array} \right).
\label{2u}
\end{equation}
As a particular example of the  Cheon-Shigehara approach \cite{cs}, the $\delta'$-interaction can 
be constructed from  the spatially symmetric configuration consisting of 
three separated $\delta$-potentials having the intensities scaled in a nonlinear way as 
the distances between the potentials tend to zero. Following this approach,  
Exner, Neidhardt and Zagrebnov \cite{enz}  have approximated the $\delta$-potentials  by regular
functions and realized rigorously the similar one-point limit in the norm resolvent topology. In particular,
they have proved that the resulting limit takes place if the distances between the 
peaks of $\delta$-like regularized potentials tend to zero
sufficiently slow relative to shrinking these potentials to the origin.
The other aspects of the $\delta'$-interaction and its approximations by local and nonlocal potentials have  been investigated, for instance,
 by Albeverio and Nizhnik \cite{an00,an06,an07,an13}, Fassari and Rinaldi \cite{fr}
 (see also references therein). The $\delta'$-interaction can be used together with
background potentials. Thus,   Albeverio, Fassari and Rinaldi \cite{afr1}  
 have rigorously defined the self-adjoint Hamiltonian of the harmonic oscillator
perturbed by an attractive $\delta'$-interaction of strength $\beta$ centered at the origin $x=0$
(the bottom of a confining parabolic potential), explicitly providing its resolvent. 
 In a subsequent  publication \cite{afr2}, their study has been extended 
for the perturbation by a triple of attractive $\delta'$-interactions using the Cheon-Shigehara approximation. 
It is worth mentioning the recent publication  \cite{g18}, where
 Golovaty has constructed a new approximation to the 
$\delta'$-interaction involving two parameters in the boundary conditions. Here the 
 connection matrix  
 \begin{equation}
\Lambda = \left(\begin{array}{cc} \theta~~~~\beta \\ 0 ~~~ \theta^{-1} \end{array} \right)
\label{3u}
\end{equation}
 describes  the two-parametric family of point interactions being the generalization of the  
$\delta'$-interaction with $\theta = 1$.

It should be emphasized that the term ``$\delta'$-interaction'' is somewhat misleading because 
the point interaction described by the $\Lambda$-matrix (\ref{2u}) does not correspond to Equation (\ref{1}) 
in which the potential part is the derivative of the Dirac delta function in the distributional sense, i.e.,
 $V(x)=\gamma \delta'(x)$ with strength  $\gamma$. Since the term $\delta'(x)\psi(x)$ is not defined for
discontinuous $\psi(x)$, Kurasov \cite{k} has developed the distribution theory based on the space
of discontinuous at  $x = 0$ test functions. Within this theory,  as a particular example, 
the point interaction that corresponds to the potential $V(x)=\gamma \delta'(x)$ is given 
by the connection matrix 
\begin{equation}
\Lambda = \left(\begin{array}{cc} \theta~~~~0 ~\\ 0 ~~~ \theta^{-1} \end{array} \right), 
\label{4u}
\end{equation}
 where $\theta = (2 +\gamma)/(2- \gamma)$,
$\gamma \in \R\setminus \{ \pm \,2 \}$. Since the term ``$\delta'$-interaction''
is reserved for the case with the connection matrix of the type (\ref{2u}), Brasche and Nizhnik \cite{bn}
suggested to refer the point interactions described by the matrices of the form (\ref{4u}) even if
the element $\theta \neq 1$ does not correspond to the delta prime potential. We will follow this 
terminology in the present paper.

The Kurasov approach has been followed in many 
applications (see, e.g., \cite{bn,gnn,ggn,l1,l2,gggm,kp,gmmn}) including more general examples.
Thus, in the context of this approach,  Gadella {\it et al} \cite{gnn} have shown that Equation (\ref{1})
with the potential  $V(x) = a \delta(x) + b \delta'(x)$, $a <0$, $b \in \R$, 
 has a bound state and calculated the energy of this state in terms of the parameters $a$ and $b$.
A new approach based on the integral form of the Schr\"{o}dinger equation (\ref{1})
has been developed by Lange \cite{l1,l2} with some revision of  Kurasov's theory. 
The potential  $V(x) = a \delta(x) + b \delta'(x)$ has also been used by Gadella and coworkers
as a perturbation of some background 
potential, such as a constant electric field and the harmonic oscillator \cite{ggn} or
the infinite square well \cite{gggm}. 
The spectrum of  a one-dimensional V-shaped quantum well perturbed by three types of 
 a point impurity as well as 
three solvable two-dimensional systems (the isotropic harmonic oscillator,
a square pyramidal potential and their combination) perturbed by a point interaction centered at the
origin has  been studied by Fassari {\it et al} in the recent papers \cite{fggn1,fggn2,f-r}.

On the other hand, as derived in the series of publications \cite{c-g,zci,tn,z10,zz14} 
for some particular cases and proved rigorously 
by Golovaty with coworkers \cite{gm,gh,g1,gh1,g2} in a general case, 
the potential $V(x) =\gamma \delta'(x)$
appears to be partially transparent at some discrete values forming a countable set 
$\{ \gamma_n \}$ in the $\gamma$-space.
The corresponding $\Lambda$-matrix is diagonal, i.e., of the form (\ref{4u}) where the element
 $\theta = \{\theta_n\}$ takes discrete values that depend on the sequence $\{ \gamma_n \}$.
Except the distribution $\delta'(x)$, which is obtained as a limit of 
regular $\delta'$-like functions,
the diagonal form of the $\Lambda$-matrix can be realized even if the squeezed limit of 
regular functions
does not exist. Beyond the ``resonance'' set $\{ \gamma_n \}$, the $\delta'$-potential is 
fully opaque
satisfying the boundary conditions of the type (\ref{2y}). However, 
this resonant-tunneling behavior 
contradicts with the $\Lambda$-matrix (\ref{4u}) where the element $\theta$ 
continuously depends on
strength $\gamma$. It is remarkable that this controversy can be resolved using 
the one-dimensional model for the heterostructure  
consisting of two or three squeezed  parallel homogeneous layers approaching to one point \cite{z17,z18aop}. Here a ``splitting'' effect of one-point interactions has been described. 

As for two-point interactions in one dimension, one should mention the recent studies 
concerning quantum 
tunneling times and the associated questions such as, for instance, the Hartman effect 
and its 
generalized version (see, e.g., \cite{clm,lmn,clmn,llmn1,llmn2} and references therein). 
Another important 
aspect regarding the application of double-point potentials is the Casimir 
effect that arises in the behavior
of the vacuum energy between two homogeneous parallel plates. 
For the interpretation of this effect,
 Mu\~{n}oz-Casta\~{n}eda and coworkers \cite{ag-am-c,am-c1,gm-c,am-c2,m-cgm,m-ckb,m-cg} 
 reformulated the theory of self-adjoint extensions of symmetric operators over
bounded domains in the framework of quantum field theory. Particularly, they 
have calculated the vacuum energy and identified which boundary conditions 
generate attractive or repulsive
Casimir forces between the plates. Bordag and Mu\~{n}oz-Casta\~{n}eda \cite{bm-c}
have calculated the quantum vacuum interaction energy between two kinks of the
sine-Gordon equation (for a review on nonlinear localized excitations including 
topological solitons see, e.g., the work \cite{ht})
 and shown that this interaction induces an attractive force between the kinks 
in parallel to the Casimir force between conducting mirrors.  A rigorous
mathematical model of real metamaterials has been suggested in \cite{n-r}.
The resonant tunneling through double-barrier scatters is still 
an active area of research for the applications to
nanotechnology. In the context of the  Cheon-F\"{u}l\"{o}p-Tsutsui approach \cite{cft,tfc}, 
the conditions for the parameter space under which the perfect resonant transmission occurs 
through two point interactions, each of which is described by four parameters, 
have been found by
Konno,  Nagasawa and  Takahashi \cite{knt1,knt2}.

 The  pioneering  studies  \cite{te,cet,ec} demonstrated that  the resonant transmission
through quantum multilayer  heterostructures of 
electronic tunnel systems are  of considerable general
interest. These structures  are not only important in micro- and nanodevices,
but their study involves a great deal of basic physics. In
recent years it has been realized that the study of the electron transmission
through heterostructures can be investigated in the zero-thickness limit
approximation materialized when their width shrinks to zero.
Within such an approximation it is possible 
to produce various point interaction models, particularly those as described above
which  admit exact closed analytic solutions.  These models are required to  provide  
relatively
simple configurations where an appropriate way of squeezing to the zero-width limit
must be compatible with the original real structure.  
Additionally,  as a rule, the nanodevices  are subject to electric fields applied externally. 
In this regard, 
 is of great interest  to produce  point interaction models with bias potentials. So far no
 models have been elaborated for such devices using one-point approximation  methods.  
 
The present paper is devoted to the investigation of planar 
heterostructures composed of
 extremely thin layers separated by small distances in the limit where
both the layer thickness and the distance between the layers  simultaneously
tend to zero. 
 The electron motion in the systems of this type is usually 
 confined in the longitudinal
direction (say, along the $x$-axis); the latter  is perpendicular to the transverse planes 
where electronic motion is free. 
The three-dimensional Schr\"{o}dinger equation of such a structure can be separated
into longitudinal and transverse parts, writing the total electron energy
as the sum of the longitudinal and transverse energies: 
$E_l+ \hbar^2 {\bf k}_t^2/2m^*$, where
$m^{*}$ is an effective electron mass and ${\bf k}_t$ the transverse wave vector; for such
additive Hamiltonian the wave function is expressed as a product, i.e. 
 $\psi  =\psi_l \psi_t$. As a result, we arrive at the reduced one-dimensional
Schr\"{o}dinger equation with respect to the longitudinal component of the
wave function $\psi_l(x)$ and the electron energy $E_l$. For brevity of notations,
in the following we omit the subscript ``$l$'' at both  $\psi_l(x)$ and $E_l$. Thus,
in the units as $\hbar^2/2m^*=1$, 
 the one-dimensional stationary Schr\"{o}dinger equation reduces to the form (\ref{1})
where $V(x)$ is a potential for electrons. 
Concerning the dimensions of the longitudinal electron position $x$,
the potential $V(x)$ and the electron energy $E$,
 in the system $\hbar^2/2m^* =1$ we have  $[x]$ = nm and $[V,E]$ = nm$^{-2}$.  
 For computations we choose $m^* =0.1 \, m_e$ and in this case, 1 eV = 2.62464 nm$^{-2}$.

%===========================================Introduction================================

\section{Transmission characteristics of multi-layered structures} 

This introductory section generalizes the approach described in \cite{lf}.
We consider  the Schr\"{o}dinger equation (\ref{1}),
where the potential $V(x)$ is an  arbitrary  piecewise function defined on 
the interval $(x_0, x_N)$ with $N$ subsets 
$(x_{i-1}, x_{i})$, $i =\overline{1,N},~N = 1, 2, \ldots$ .
Each $ V_i(x)$ is a real bounded function  
defined on this interval, so that we have
the set of functions: $V_1(x), \ldots , V_{N}(x)$. Next, we express
the transmission matrix in terms of the interface values of the linearly
independent solutions of the Schr\"{o}dinger equation. 

The solution of the Schr\"{o}dinger equation across the interval
$(x_{i-1}, x_{i})$, $\psi_i(x)$,  will be given as
\begin{equation}
\psi_i(x)= C_i^{(1)}u_i(x) + C_i^{(2)}v_i(x),~~\overline{1,N},
\label{2}
\end{equation}
where $u_i(x)$ and $v_i(x)$ are linearly independent solutions on the interval
$(x_{i-1}, x_i)$. At the interface $x_i$, $i= \overline{1, N-1}$, the
particle conservation requires the continuity of the wave function $\psi(x)$, 
while the momentum 
conservation demands the continuity of the first derivative of the
wave function $\psi'(x)$ resulting in  the equations
\begin{equation}
\psi_i(x_i) = \psi_{i+1}(x_i), ~~\psi'_i(x_i) = \psi'_{i+1}(x_i),
~~i = \overline{1,  N-1},
\label{3}
\end{equation}
where the prime denotes first derivative with respect to $x$.

\subsection{Transmission matrix}

Using Equation (\ref{2}), the boundary conditions 
(\ref{3}) can be realized as a system of two
linear equations with two unknowns such that
\begin{equation}
{\bf M}_i(x_i){\bf C}_i = {\bf M}_{i+1}(x_{i}){\bf C}_{i+1},~~~i = 
\overline{1, N-1},
~~N \ge 2,
\label{4}
\end{equation}
where 
\begin{equation}
{\bf C}_i := \mbox{col}\left(C_i^{(1)}, C_i^{(2)}\right) = 
\left(\begin{array}{cc}
C_i^{(1)} \\  C_i^{(2)} \end{array} \right)~~\mbox{and}~~
{\bf M}_i(x) :=  \left( \begin{array}{cc} u_i(x)~~v_i(x) \\
u_i'(x) ~~v'_{i}(x) \end{array} \right)
\label{5}
\end{equation}
are Wronskian matrices.
Next, using Equations (\ref{4}),  one can connect the column vectors 
${\bf C}_1$ and ${\bf C}_N$  as follows
\begin{eqnarray}
{\bf C}_N &=& {\bf M}^{-1}_N(x_{N-1}){\bf M}_{N-1}(x_{N-1})
{\bf M}^{-1}_{N-1}(x_{N-2})
\ldots {\bf M}_2(x_{2}){\bf M}^{-1}_2(x_1){\bf M}_1(x_1){\bf C}_1 \nonumber \\
&=& {\bf M}^{-1}_N(x_{N-1}) \Lambda_{N-1}(x_{N-2}, x_{N-1}) \ldots 
\Lambda_{2}(x_1, x_2){\bf M}_1(x_1){\bf C}_1 ,~~N \ge 2,
\label{6}
\end{eqnarray}
where we have introduced the following matrices: 
\begin{equation}
\Lambda_i(x_{i-1}, x_i) := {\bf M}_{i}(x_{i}){\bf M}^{-1}_{i}(x_{i-1}),
~~~i =\overline{2, N-1},~~N \ge 3.
\label{7}
\end{equation}
Here each matrix $\Lambda_i(x_{i-1}, x_i)$ connects the boundary values of
the corresponding Wronskian matrix ${\bf M}_i(x)$ at $x =x_{i-1}$ and $x =x_{i}$. 
Yet, it is not obvious that the matrices $\Lambda_i$'s are
transmission matrices connecting the boundary conditions imposed on the wave functions
$\psi_i(x)$ at  $x =x_{i-1}$ and $x =x_{i}$. To prove this fact, we compute the right-hand
matrix product of (\ref{7}) and obtain
\begin{equation}
\Lambda_i(x_{i-1}, x_i) = 
 \left( \begin{array}{cc} \lambda_{i,11}~~\lambda_{i,12} \\
 \lambda_{i,21}~~\lambda_{i,22} \end{array} \right) ,
\label{8}
\end{equation}
where
\begin{eqnarray}
\lambda_{i,11}(x_{i-1}, x_i) &=& \left[ u_i(x_{i})v'_i(x_{i-1}) 
- u'_i(x_{i-1})v_i(x_i)\right]/W_i\,, \nonumber \\
\lambda_{i,12}(x_{i-1}, x_i) &= & \left[ u_i(x_{i-1})v_i(x_i) 
- u_i(x_i)v_i(x_{i-1})\right]/W_i\,, \nonumber \\
\lambda_{i,21}(x_{i-1}, x_i) &= & \left[ u'_i(x_i)v'_i(x_{i-1}) 
- u'_i(x_{i-1})v'_i(x_i)\right]/W_i\,, \nonumber \\
\lambda_{i,22}(x_{i-1}, x_i) &=& \left[ u_i(x_{i-1})v'_i(x_i) 
- u'_i(x_i)v_i(x_{i-1})\right]/W_i\,,
\label{9}
\end{eqnarray}
with the Wronskian
\begin{equation}
W_i  =  W_i(x_{i-1})= u_i(x_{i-1})v'_i(x_{i-1}) - u'_i(x_{i-1})v_i(x_{i-1})
\label{10}
\end{equation}
computed at $x=x_{i-1}$, which does not depend on $x$ on the interval
$( x_{i-1},  x_i)$. 
Using Equations (\ref{9}) and (\ref{10}), one can check that
$\det\Lambda_i = 1.$

  There is an infinite number of the linearly independent solutions $u_i(x)$ and 
 $v_i(x)$. The representation of the $\Lambda_i$-matrix elements can be 
 simplified if we choose these solutions satisfying the initial conditions:
 \begin{equation}
 u_i(x_{i-1})=1,~~u_i'(x_{i-1})=0,~~v_i(x_{i-1})=0,~~v_i'(x_{i-1})=1.
 \label{12}
 \end{equation}
 Inserting thus these conditions into Equations (\ref{9}) and (\ref{10}), we get
 that $W_i =1$ and, as a result,
\begin{equation} 
\Lambda_i(x_{i-1}, x_i) = 
 \left( \begin{array}{cc} u_i(x_i)~~~~ v_i(x_i) \\
 u'_{i}(x_i)~~~~v'_{i}(x_i) \end{array} \right) .
\label{13}
\end{equation}
 
The next step is to compute the product 
$\Lambda_i(x_{i-1}, x_i) \mbox{col}\left(\psi_{i}(x_{i-1}), \psi'_i(x_{i-1})\right)$.
This computation immediately 
results in $\mbox{col}\left(\psi_{i}(x_{i}), \psi'_i(x_{i})\right)$, so that we have 
the matrix relation
\begin{equation}
 \left(\begin{array}{cc}
\psi_i(x_i) \\ \psi'_i(x_i) \end{array} \right)= \Lambda_i(x_{i-1}, x_i)
\left(\begin{array}{cc}
\psi_i(x_{i-1}) \\ \psi'_i(x_{i-1}) \end{array} \right),
\label{14}
\end{equation}
confirming that Equation (\ref{7}) indeed defines the transmission matrix 
$\Lambda_i(x_{i-1}, x_i)$
expressed in terms of the matrices ${\bf M}_{i}(x_{i-1})$ 
and ${\bf M}_{i}(x_{i})$.
Thus, each transmission matrix $\Lambda_i(x_{i-1}, x_i)$ connects the boundary 
conditions at $x =x_{i-1}$ and $x=x_i$.

Equation (\ref{6}) that connects the column vectors ${\bf C}_1$ and ${\bf C}_N$
can be transformed to the equation connecting the boundary conditions at $x=x_0$
and $x=x_N$. To this end, we define the lateral transmission matrices 
$\Lambda_i(x_{i-1},x_i)$ with $i =0, N$. Thus, on one side, one can write 
\begin{equation}
{\bf M}_1(x_1){\bf C}_1 = \left(\begin{array}{cc}
\psi_1(x_1) \\ \psi'_1(x_1) \end{array} \right) = \Lambda_1(x_0, x_1)
\left(\begin{array}{cc}
\psi_1(x_0) \\ \psi'_1(x_0) \end{array} \right).
\label{15}
\end{equation}
On the other hand, multiplying from the left Equation (\ref{6}) by 
${\bf M}_N(x_N)$ and using that
\begin{equation}
{\bf M}_N(x_N){\bf C}_N = \left(\begin{array}{cc}
\psi_N(x_N) \\ \psi'_N(x_N) \end{array} \right), 
\label{16}
\end{equation}
one finds the relation that connects the boundary conditions at $x=x_0$ and $x=x_N$:
\begin{equation}
\left(\begin{array}{cc}
\psi_N(x_N) \\ \psi'_N(x_N) \end{array} \right)= \Lambda(x_0, x_N)
\left(\begin{array}{cc}
\psi_1(x_0) \\ \psi'_1(x_0) \end{array} \right)
\label{17}
\end{equation}
with 
\begin{equation}
\Lambda(x_0, x_N) = \Lambda_N(x_{N-1}, x_N)\ldots \Lambda_1(x_0, x_1).
\label{18}
\end{equation} 
Thus, the transmission matrix for each layer defined on the interval 
$(x_{i-1}, x_i)$ can be computed through the solutions
$u_i(x)$ and $v_i(x)$ and their derivatives taken at the boundaries $x=x_{i-1}$ 
and $x=x_i$, resulting in the elements given by Equations (\ref{9}) and (\ref{10}).

\subsection{Reflection-transmission coefficients}

Consider now the solutions outside the interval $(x_0,x_N)$. In the
region $x < x_0$  and $ x > x_N$ where the potential is a constant,
the wave function is the well-known free particle  solution of
the Schr\"{o}dinger equation (\ref{1}) as follows
\begin{equation}
\psi_0(x) = A_1 \exp[ik_L(x- x_0 )] + A_2 \exp[ -ik_L(x- x_0)]
\label{19}
\end{equation}
for $x < x_0$ and 
\begin{equation}
\psi_{N+1}(x) = B_1 \exp[ik_R(x- x_N)] + B_2 \exp[ -ik_R(x- x_N)]
\label{20}
\end{equation}
for $x > x_N$, where $k_L := \sqrt{E - V_L}$ and $k_R := \sqrt{E - V_R}\,$.
Then the continuity of the boundary conditions at $x = x_0 $ and $x = x_N$ leads
to the following equations:
\begin{eqnarray}
&& \psi_0(x_0)=\psi_1(x_0),~~\psi'_0(x_0)=\psi'_1(x_0),\nonumber \\
&& \psi_N(x_N)=\psi_{N+1}(x_N),~~\psi'_N(x_N)=\psi'_{N+1}(x_N) ,
\label{21}
\end{eqnarray}
which can be represented in the matrix form as follows
\begin{equation}
{\bf M}_L {\bf A} ={\bf M}_1(x_0){\bf C}_1, ~~~
{\bf M}_N(x_N){\bf C}_N ={\bf M}_R {\bf B},
\label{22}
\end{equation}
where ${\bf A} := \mbox{col}(A_1, A_2),$ ${\bf B} := \mbox{col}(B_1, B_2)$ and
\begin{equation}
{\bf M}_L := \left(\begin{array}{cc}
1~~~~~~~~ 1 \\ ik_L ~ -ik_L \end{array} \right)\!,~~~
{\bf M}_R := \left(\begin{array}{cc}
1~~ ~~~~~~1 \\ ik_R ~ -ik_R \end{array} \right)\!.
\label{23}
\end{equation}
Using these matrix equations in Equation (\ref{6}), 
we obtain the following basic equation,
which allows us to represent the reflection-transmission coefficients through 
the elements (\ref{9}) of the transmission matrix $\Lambda(x_0,x_N)$:
\begin{equation}
\Lambda(x_0, x_N){\bf M}_L {\bf A}= {\bf M}_R {\bf B} .
\label{24}
\end{equation}

Thus,  if there is no incidental particle coming from the right, 
one can  set 
\begin{equation}
A_1 =1,~~A_2 =  R_L,~~B_1 =  T_L,~~B_2 =0,
\label{c24}
\end{equation}
so that in Equation (\ref{24}) we have ${\bf A}= \mbox{col}(1,R_L)$ and 
${\bf B} = \mbox{col}(T_L,0)$. 
Similarly, if there is no incidental particle from the left, we put 
\begin{equation}
A_1 = 0, ~~A_2 = T_R, ~~B_1 =R_R,~~B_2 = 1,
\label{c25}
\end{equation}
hence  ${\bf A}= \mbox{col}(0,T_R)$ and ${\bf B} = \mbox{col}(R_R,1)$ in (\ref{24}). 
Then Equation (\ref{23}) becomes a set
of two linear equations with respect to the pair $\{R_L, T_L\}$ or 
$\{R_R, T_R\}$. Solving these equations
and using the relation $\lambda_{11}\lambda_{22} - \lambda_{12}\lambda_{21} =1$,
we find
\begin{equation}
R_L = -\,{ p +{\rm i}q \over D},~~T_L = { 2k_L/k_R \over D},~~
R_R = {p-{\rm i}q \over D},~~T_R= {2\over D},
\label{25}
\end{equation}
where
\begin{equation}
p:= \lambda_{11} -(k_L/k_R)\lambda_{22}\,,~~~ 
q:= k_L \lambda_{12} + k_R^{-1}\lambda_{21}
\label{28}
\end{equation}
and 
\begin{equation}
D := \lambda_{11} + (k_L/k_R)\lambda_{22}- {\rm i}(k_L \lambda_{12}- k_R^{-1}\lambda_{21}).
\label{26}
\end{equation}

The current $j(x) =({\rm i}/2) (\psi \partial_x\psi^* -\psi^* \partial_x\psi )$
has to be conserved across the transition region $x_0 \le x \le x_N\,$.
Using the definition of the reflection-transmission coefficients given above, we find
the left-to-right current $j_L(x_0)=  k_L (1- |R_L|^2),$ $j_L(x_N)= k_R |T_L|^2$
and the right-to-left current $j_R(x_0)=- \, k_L |T_R|^2,$ $j_R(x_N)= 
-\, k_R(1 - |R_R|^2)$.
From the equations $j_{L,R}(x_0) = j_{L,R}(x_N)$ we obtain the conservation law
for both the directions of the current: ${\cal R}_{L,R} + {\cal T}_{L,R}=1$,
where
\begin{equation}
{\cal R}_L := |R_L|^2,~~{\cal T}_L := (k_R/k_L)|T_L|^2, ~~{\cal R}_R := 
|R_R|^2,~~{\cal T}_R := (k_L/k_R)|T_R|^2.
\label{c27}
\end{equation}
One can derive that $|D|^2 = 4k_L/k_R + p^2 +q^2$ and, as a result,
the reflection-transmission amplitudes can be represented in the form
\begin{equation}
{\cal R}_{L,R} = {p^2 +q^2 \over 4k_L/k_R + p^2 +q^2}\,, ~~~
{\cal T}_{L,R}  = { 4k_L/k_R \over 4k_L/k_R + p^2 +q^2}\,.
\label{27}
\end{equation}
In its turn, the scattering matrix can also 
be represented in terms of the elements of the
transmission matrix $\Lambda$. Indeed, due to Equations (\ref{25}) and (\ref{c27}),
this representation reads
\begin{equation}
S = \left( \begin{array}{cc} ~~~~~R_L~~~~~~~~~\sqrt{k_L/k_R}\,T_R \\
\sqrt{k_R/k_L}\,T_L ~~~~~~~~~R_R~~~~~ \end{array} \right)= 
{1 \over D} \left(\begin{array}{lr} -\,p-{\rm i}q~~~~~~2 \sqrt{k_L/k_R} \\
2 \sqrt{k_L/k_R}~~~~~~~~ p-{\rm i}q \end{array} \right), 
\label{c28}
\end{equation}
where $p$, $q$ and $D$ are defined by Equations (\ref{28}) and (\ref{26}).

\section{Schr\"{o}dinger equation and transmission matrix for the layer with  
a linear potential profile}

Consider now  the particular case of a linear potential profile 
 for the layer defined on the interval $(x_{i-1}, x_i)$. In this case
the solutions $u_i(x)$ and $v_i(x)$  and thus the transmission matrix 
$\Lambda_i(x_{i-1},x_i)$ can be written explicitly.
The Schr\"{o}dinger equation (\ref{1}) for the $i$th layer, $i =\overline{1,N}$,
 can be rewritten as
\begin{equation}
- \psi_i''(x) +V_i(x)\psi_i(x) = E \psi_i(x) ,
\label{29}
\end{equation}
  where the potential $V_i(x)$ is a 
linear function defined on the interval $x_{i-1} < x < x_i$
 of length  $l_i := x_{i} - x_{i-i}$, i.e., 
\begin{equation}
V_i(x) = \eta_i (x-x_i) +V_i(x_i), ~~\eta_i :={ V_i(x_i) - V_i(x_{i-1})
\over  l_i} \, .
\label{30}
\end{equation}
These equations can be transformed to the  Airy equation 
\begin{equation}
{ d^2\psi_i(z_i) \over dz_i^2} - z_i \psi_i(z_i)=0,
\label{31}
\end{equation}
 by setting  $ z_i(x) = \sigma_i(x- s_i)$,
where the constants $\sigma_i$ and $s_i$ are given by 
\begin{equation}
\sigma_i = \eta_i^{1/3} ,~~~ s_i = x_i + \eta^{-1}_i [E - V_i(x_i)].
\label{32}
\end{equation}

According to the general expressions (\ref{9}), we use the Airy functions
 of the first and  the second order as linearly independent
 solutions to Equation (\ref{31}), setting
 $u_i(x) = Ai(z_i(x))$ and $v_i(x) = Bi(z_i(x))$. 
 On the interval $- \infty < z_i < \infty$, these solutions are real-valued.
 The interface (boundary)
 values of the (dimensionless) 
 function $z_i(x)$ at the edges of the $i$th layer, to be used in
 Equations (\ref{9}) and (\ref{10}), are 
 \begin{equation}
z_{i,i-1} := z_i(x) |_{x= x_{i-1}} = -\, \eta_i^{- 2/3}k^2_{i,i-1}\,,~~~
z_{i,i} := z_i(x) |_{x= x_{i}} = -\, \eta_i^{- 2/3}k^2_{i,i}\,,
\label{33}
\end{equation}
where
\begin{equation}
k_{i,i-1} := \sqrt{E- V_{i,i-1}}\, , ~k_{i,i}:= \sqrt{E- V_{i,i}}\, ,~
V_{i,i-1} := V_i(x_{i-1}),~V_{i,i} := V_i(x_{i}). \nonumber \\
\label{34}
\end{equation}
 The Wronskian with respect to the variable $z$
  is $W\{ Ai(z), Bi(z) \} =1/\pi$, 
 therefore with respect to $x$, it is $W\{ Ai(z_i(x)), Bi(z_i(x)) \} =\sigma_i/\pi$.
 Then the elements of the $\Lambda_i$-matrix are
\begin{eqnarray}
\lambda_{i,11}(x_{i-1}, x_i) &=& \pi \left[ Ai(z_{i,i})Bi'(z_{i,i-1}) 
- Ai'(z_{i,i-1})Bi(z_{i,i})\right], \nonumber \\
\lambda_{i,12}(x_{i-1}, x_i) &=& (\pi/\sigma_i) \left[ Ai(z_{i,i-1})Bi(z_{i,i}) 
- Ai(z_{i,i})Bi(z_{i,i-1})\right], \nonumber \\
\lambda_{i,21}(x_{i-1}, x_i) &=& \sigma_i \pi \left[ Ai'(z_{i,i})Bi'(z_{i,i-1}) 
- Ai'(z_{i,i-1})Bi'(z_{i,i})\right], \nonumber \\
\lambda_{i,22}(x_{i-1}, x_i) &=& \pi \left[ Ai(z_{i,i-1})Bi'(z_{i,i}) 
- Ai'(z_{i,i})Bi(z_{i,i-1})\right],
\label{35}
\end{eqnarray}
where the prime denotes the differentiation with respect to $z$.

In the $\eta_i \to 0$ limit as $V_i(x_{i-1}) \to V_i(x_i)$, we obtain
\begin{equation}
z_i(x) \to - \, \sigma_i s_i =\sigma_i
 \left[- x_i - {E -V_i(x_i) \over \sigma_i^3} \right]
\to \sigma_i^{-2} \left[  V_i(x_i) -E \right], 
\label{36}
\end{equation}
yielding  Equation (\ref{29}) with a constant profile $V_i(x) \equiv V_i$.
In this limit case,  one can choose the linearly independent solutions
to Equation (\ref{29}) as
\begin{equation}
u_i(x)=\cos[k_i(x-x_{i-1})],~v_i(x)=k_i^{- 1}\sin[k_i(x-x_{i-1})],
~k_i:= \sqrt{k^2 -V_i}\, ,
\label{37}
\end{equation}
satisfying the initial conditions (\ref{12}). Therefore, 
due to Equations (\ref{13}) and (\ref{37}), the $\Lambda_i$-matrix becomes 
 \begin{equation}
\Lambda_i(x_{i-1}, x_i) = \left(\begin{array}{cc}
~~~ ~\cos(k_il_i)~~~~ k_i^{-1}\sin(k_il_i) \\ 
-\, k_i\sin(k_il_i)~~~~ \cos(k_il_i) 
\end{array} \right) .
\label{38}
\end{equation}

\section{Asymptotic representations of the single-layer transmission matrix}  
  
  Similarly to the previous section, here we also focus on one of the layers 
  and for brevity of notations we replace for while in the above expressions
  the subscripts 
  $\{i,i-1\}$ and $\{i,i\}$ by ``0'' and ``1'', respectively. Then, according to
  Equations (\ref{33}) and (\ref{34}), we write
 \begin{equation}
z_0  =- \left({l \over V_1 -V_0}\right)^{\!\!2/3}k_0^2, ~
z_1 = - \left({l \over V_1 -V_0}\right)^{\!\!2/3}k_1^2,~\sigma = 
\left( {V_1 -V_0 \over l} \right)^{\!\!1/3},
\label{39}
\end{equation} 
 where we have replaced $V_{i,i-1}$, $V_{i,i}$, $k_{i,i-1}$, $k_{i,i}$ by 
  $V_0$, $V_1$, $k_0$, $k_1$, respectively. Using next
  the two asymptotic expressions for the Airy functions and their derivatives 
  known in the limit as  $z \to 0$ and $z \to \pm \, \infty$, below we will
  derive the corresponding asymptotic representations  of the elements 
  (\ref{35}) in the two limits as  (i) $z_0, z_1 \to 0$ and (ii) 
  $z_0, z_1 \to \pm \, \infty$. It is reasonable to assume that everywhere 
  $z_0$ and $z_1$ are of the same sign. We  omit for a while the subscript
  ``$i$'' for the matrix $\Lambda_i$ and its elements.

 \subsection{Asymptotic representation of the $\Lambda$-matrix 
in the limit as  $z_{0}, z_1 \to 0$}

For the  $z_0, z_1 \to 0$ limit to be carried out  in Equations (\ref{35}), 
 one can use the series representation of the Airy functions and their 
first derivatives in the neighborhood of the origin $z =0$.  It is sufficient to 
explore only the two first  terms:
\begin{eqnarray}
Ai(z) & \to & {1 \over 3^{2/3} \Gamma(2/3)} - {z \over 3^{1/3} \Gamma(1/3)} +\ldots ,
\nonumber \\
Ai'(z) & \to & - \, {1 \over 3^{1/3} \Gamma(1/3)} + {z^2 \over 2 
\cdot 3^{2/3} \Gamma(2/3)}
+ \ldots , \nonumber \\
Bi(z) & \to & {1 \over 3^{1/6} \Gamma(2/3)} + {3^{1/6} z \over  \Gamma(1/3)} + 
\ldots , \nonumber \\
Bi'(z) &\to & {3^{1/6} \over \Gamma(1/3)} + { z^2 \over 2 \cdot 3^{1/6}
 \Gamma(2/3)} + \ldots . 
 \label{40}
\end{eqnarray}
As a result of applying these expansion formulae to Equations (\ref{35})
and using Euler's reflection formula for the gamma function, 
$\Gamma(1-z)\Gamma(z)= \pi/\sin(\pi z),~z \notin \Z$, we get
the following asymptotic representation of the $\Lambda$-matrix elements: 
\begin{eqnarray}
\lambda_{11} & \to & 1 -  z_0^2 z_1/2 , ~~ \lambda_{22} \to 1 - z_0 z_1^2/2,
\nonumber \\
  \lambda_{12} &\to & {z_1 - z_0 \over \sigma} =l,~~
\lambda_{21} \to {\sigma \over 2} (z_1^2 - z_0^2) = -\, { l \over 2} (k_0^2 + k_1^2)
\label{41}
\end{eqnarray}
as $z_0, z_1 \to 0$.

\subsection{Asymptotic representation of the $\Lambda$-matrix in the limit
as $z_0, z_1 \to \pm \, \infty$ }

In the limit as $z \to - \, \infty$, 
for the Airy functions and their derivatives we have the following asymptotics:
\begin{equation}
Ai(z) \to { \sin\!\left[ {2 \over 3}(- z)^{3/2} + \pi /4 \right] \over \sqrt{\pi}
(-z)^{1/4} }, ~~ 
Bi(z) \to { \cos\!\left[ {2 \over 3}(- z)^{3/2} + \pi /4 \right] \over \sqrt{\pi}
(-z)^{1/4} },
\label{42}
\end{equation}
\begin{eqnarray}
Ai'(z) &\to & { \frac14 (-z)^{-3/4} \sin\!\left[ {2 \over 3}(- z)^{3/2} + \pi /4 \right] 
- (- z)^{3/4} \cos\!\left[ {2 \over 3}(- z)^{3/2} + \pi /4 \right] \over \sqrt{\pi}
(-z)^{1/2} }, ~~ \nonumber \\
Bi'(z) &\to & { \frac14 (-z)^{-3/4} \cos\!\left[ {2 \over 3}(- z)^{3/2} + \pi /4 \right] +
(- z)^{3/4} \sin\!\left[ {2 \over 3}(- z)^{3/2} + \pi /4 \right]  
\over \sqrt{\pi} (-z)^{1/2} }   . \nonumber \\
\label{43}
\end{eqnarray}
Using this asymptotic representation in Equations (\ref{35}) as 
 $z_0, z_1 \to -\, \infty$, we obtain
\begin{eqnarray}
\lambda_{11} &\to & (- z_0)^{1/4}( -z_1)^{-1/4}\cos\!\chi_- 
- (4 z_0)^{-1}(- z_0)^{-1/4} (- z_1)^{-1/4}\sin\!\chi_- \, , 
\nonumber \\
\lambda_{12} &\to & -\, \sigma^{-1} (- z_0)^{-1/4} (- z_1)^{-1/4}\sin\!\chi_- \, , 
\nonumber \\
\lambda_{21} &\to & \sigma\, (- z_0)^{- 1/2} (- z_1)^{-1/2} \nonumber \\
&&  \times  \left\{ \left[  (- z_0)^{3/4} (- z_1)^{3/4} + 4^{-2} 
 (- z_0)^{- 3/4} (- z_1)^{- 3/4} 
 \right]  \sin\!\chi_- \right. \nonumber \\
 & & + \left. 4^{-1} \left[   (- z_0)^{3/4} (- z_1 )^{- 3/4} 
 -   (- z_1)^{3/4} (- z_0)^{- 3/4}\right] \cos\!\chi_- \right\}   , \nonumber \\
\lambda_{22} &\to & (- z_1)^{1/4} (- z_0)^{- 1/4}\cos\!\chi_- 
+ (4 z_1)^{-1}(- z_0)^{- 1/4} (- z_1)^{-1/4}\sin\!\chi_-\, , ~~~~~~
\label{44}
\end{eqnarray}
where 
\begin{equation}
\chi_- := \frac23 \! \left[ (- z_1)^{3/2} - (- z_0)^{3/2}\, \right].
\label{45}
\end{equation}
One can check that $\det\Lambda  =1$.  According to Equations (\ref{39}), this 
representation corresponds to a well ($V_{j} < 0$, $j=0,1$). However, 
these formulae can be ``continued'' to positive values of $z_0$ and $z_1$
that correspond to a barrier with $E < V_{j}$. 
To prove this, we  use the asymptotic representation of the Airy functions and 
their derivatives in the limit as  $z_0, z_1 \to +\, \infty$:
\begin{equation}
Ai(z) \to {{\rm e}^{-\, {2 \over 3} z^{3/2}} \over ~~~
2 \sqrt{\pi} z^{1/4}}\, ,~~~
Bi(z) \to {{\rm e}^{ {2 \over 3} z^{3/2}} \over 
 \sqrt{\pi} z^{1/4}}\, ,
\label{46}
\end{equation}
\begin{equation}
Ai'(z) \to -\, {z^{3/4} + {1 \over 4} z^{-3/4}
 \over 2 \sqrt{\pi} z^{1/2} } \, 
   {\rm e}^{-\, {2 \over 3} z^{3/2} }, ~~
Bi'(z) \to  {z^{3/4} - {1 \over 4} z^{-3/4}
 \over  \sqrt{\pi} z^{1/2} }  \,
   {\rm e}^{ {2 \over 3} z^{3/2} } 
\label{47}
\end{equation}
and, as a result, we find
\begin{eqnarray}
\lambda_{11} &\to & (z_0/z_1)^{1/4}\cosh\!\chi_+ 
+ (4 z_0)^{-1}(z_0 z_1)^{-1/4}\sinh\!\chi_+ \, , \nonumber \\
\lambda_{12} &\to &  \sigma^{-1} (z_0 z_1)^{-1/4}\sinh\!\chi_+ \, , 
\nonumber \\
\lambda_{21} &\to & \sigma \, (z_0 z_1)^{-1/2}
 \left\{ \left[  (z_0 z_1)^{3/4} - 4^{-2} (z_0 z_1)^{- 3/4} 
 \right]  \sinh\!\chi_+ \right. \nonumber \\
 & & ~~~~~ + \left. 4^{-1} \left[   ( z_1 / z_0 )^{3/4} 
 -   ( z_0 / z_1)^{3/4}\right] \cosh\!\chi_+ \right\}   , \nonumber \\
\lambda_{22} &\to & (z_1/z_0)^{1/4}\cosh\!\chi_+ 
- (4 z_1)^{-1}(z_0 z_1)^{-1/4}\sinh\!\chi_+\, , 
\label{48}
\end{eqnarray}
where $z_0$ and $z_1$ are positive and
\begin{equation}
 \chi_+ := \frac23 \! \left(z_1^{3/2} - z_0^{3/2} \right).
\label{49}
\end{equation}
Similarly, for the elements (\ref{48}) one can also check that $\det\Lambda =1$.
In fact, Equations (\ref{48}) with (\ref{49}) appear to coincide with 
Equations (\ref{44}) 
and (\ref{45}) if we assume that in the latter equations 
$z_0$ and $z_1$ are positive. To show this, we note that 
$(-z)^{3/2}= {\rm i}^3 z^{3/2} = - {\rm i} z^{3/2}$ and, as a result, 
we get the relation $\chi_- = - {\rm i}\chi_+$ for positive $z_0$ and $z_1$ in both
Equations (\ref{45}) and (\ref{49}). Next,
the elements (\ref{48}) are obtained from the representation (\ref{44})
if we note that  $ (-z_0)^{1/4}(-z_1)^{1/4}= {\rm i}(z_0 z_1)^{1/4}$,
$(-z_0)^{1/2}(-z_1)^{1/2}= - (z_0 z_1)^{1/2}$ and 
$ (-z_0)^{3/4}(-z_1)^{3/4}= - {\rm i}(z_0 z_1)^{3/4}$. 
Therefore in the following 
it is sufficient to consider only the representation given by 
Equations (\ref{44}) and (\ref{45}), being  valid for both 
 negative and positive $z_0$ and $z_1$. 

Using the explicit values for $z_0$ and $z_1$ given by Equations (\ref{39}),
 the expression (\ref{45}) for  $\chi_-$ can be transformed to 
 \begin{equation}
 \chi_- = \mbox{sgn}(V_0 - V_1) \, k_{1,0}\, l,
 \label{50}
 \end{equation}
 where
\begin{equation}
k_{1,0} := { 2( k_0^2 + k_1^2 + k_0k_1)
\over  3(k_0 + k_1) },~~~k_{j} := \sqrt{E-V_{j}} \,~~j=0,1.
\label{51}
\end{equation}
 Inserting next
the expressions (\ref{39}) and (\ref{50}) into Equations (\ref{44}), one can write 
the  elements of the $\Lambda$-matrix in terms of $k_0$ and $k_1$ as follows 
\begin{eqnarray}
\lambda_{11} & \to & \left( {k_0 \over k_1 }\right)^{\!1/2}\!
\cos(\kappa l) + {k_1^2 - k_0^2 \over 4l } k_0^{- 5/2} k_1^{- 1/2} \sin(k_{1,0}\, l) ,
\nonumber \\
\lambda_{12} & \to & k_0^{- 1/2} k_1^{- 1/2}\sin(k_{1,0} \,l), \nonumber \\
\lambda_{21} & \to &  { 3 (k_0^2 -k_1^2)^2 k_{1,0} \over 8l k_0^{ 5/2} k_1^{5/2} }
 \cos(k_{1,0} \,l)  \nonumber \\
&& -\, k_0^{1/2} k_1^{1/2} \! \left[1 + 
\left( { k_0^2 - k_1^2 \over 4l}\right)^{\!\!2}
k_0^{- 3} k_1^{- 3}\right]\! \sin(k_{1,0} \,l) , \nonumber \\
\lambda_{22} & \to &  \left( {k_1 \over k_0}\right)^{\!1/2}\!\cos(\kappa l)
+ {k_0^2 - k_1^2 \over 4l} k_0^{-1 /2} k_1^{- 5/2} \sin(k_{1,0}\, l) ,
\label{52}
\end{eqnarray}
where $k_{1,0}$ is defined by Equation (\ref{51}). One can check that 
the matrix elements (\ref{52}) together with the argument (\ref{51})
satisfy the condition $\det\Lambda=1$. 
Note that the only restriction for the 
existence of the representation (\ref{52}) are the asymptotics 
$z_0, z_1 \to \pm \, \infty$. Both $k_0$
and $k_1$ are either real-valued or imaginary. In the particular case 
$V_1 = V_0$ ($k_1 = k_0$), Equations (\ref{51}) and (\ref{52}) reduce to 
the matrix representation (\ref{38}).

\section{ Realization of  point interactions in the zero-thickness limit for one   
layer}

Keeping in the following the same notations with respect to 
the subscripts ``0'' and ``1'', 
let us consider the linear potential (\ref{30}) rewritten as
\begin{equation}
V(x)= V_0 + {V_1 - V_0 \over l}\, x,~~V_0,\, V_1 \in \R,
\label{53}
\end{equation}
 on the interval $0 < x < l$, where $V_0$ and $V_1$ are the potential 
values at the left and right edges of the layer with width $l$.
Consider first the case when this potential is constant, i.e., 
$V_0 = V_1$. A point interaction can be realized in the limit as the 
layer thickness $l \to 0$, whereas $V_0 \to \pm \, \infty$. 
To this end, one can use the 
parametrization of the potential $V(x) \equiv V_0$
introducing a dimensionless parameter $\varepsilon >0$ that controls 
the shrinking of the layer to zero width as $\varepsilon \to 0$.
It is natural to consider the power parametrization setting
\begin{equation}
V_0 = a \, \varepsilon^{- \mu}, ~~l =\varepsilon d,~~a \in \R, ~~
\mu,\, d >0.
\label{54}
\end{equation}
%-----------------------------fig1--------------------------------------
\begin{figure}
\centerline{\includegraphics[width=0.5\textwidth]{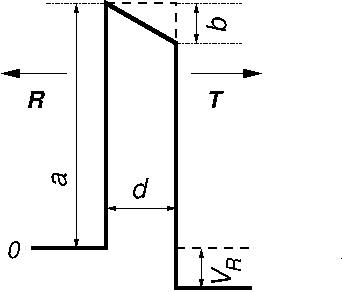}}
%\vspace{0.1pt}
\caption{ Schematics of  one-layer potential (\ref{53}) tilted by 
difference $V_1 -V_0 $ (solid line) with notations
given in (\ref{57}) at $\varepsilon =1$: $V_1 - V_0 = b = V_R$. The dashed line
represents potential with $b =0$.  }
\label{fig1}
\end{figure}
%-------------------------fig1------------------------------------ 

In the squeezed limit (as $\varepsilon \to 0$), a one-parameter family  
of point interactions at $x=0$ is realized. It is determined by 
the power $\mu \in (0, \infty)$: the transmission is perfect for 
$\mu \in (0,1)$, at $\mu =1$ the potential takes the form of
 Dirac's delta function $\alpha \delta(x)$ with the transmission matrix (\ref{1u}),
where $\alpha = ad $ is the strength of the $\delta$-interaction, and for
$\mu \in (1, \infty)$ the interaction acts as a fully reflecting wall
satisfying the Dirichlet boundary condition $\psi(\pm 0)=0$ for the wave
function $\psi(x)$.

In the case when the difference $V_1 - V_0 $ is non-zero, as shown in 
{\bf Figure \ref{fig1}},  
one deals with two potential values $V_0$ and $V_1$ at the layer edges
that must tend to infinity in the zero-thickness limit. 
Both the potential values $V_0$ and $V_1$ are supposed to be of the same sign.
In general,
the rate of this divergence to infinity
 can differ and therefore the parametrization
of the potential (\ref{53}) should involve two parameters. We introduce 
the two powers $\mu$ and $\nu$, where the parameters $\mu$ and $\nu$ describe
 how rapidly the potential $V_0$  at the left layer edge and  the difference 
$V_1 -V_0$ tend (escape) to infinity as $\varepsilon \to 0$, respectively. 
The particular case when this difference is 
a constant not depending on $\varepsilon$ can also be included. 
 Thus, we set  
\begin{equation}
V_0 = a \varepsilon^{ - \mu},~V_1 = V_0 +  b \, \varepsilon^{- \nu}, ~
0< \mu < \infty,~  0 \le \nu \le \mu ,~
  a,\, b \in \R,~l = \varepsilon \, d,~~~~~
\label{57}
\end{equation}
  including the following two situations in the squeezed limit: 
  (i) $V_1 -V_0 $ is constant ($\nu =0$)
   and (ii) the ``escaping-to-infinity'' 
    rate of $V_1 -V_0$ does not exceed the rate of $V_0$    ($\nu \le \mu$).  
In  the electronics domain the difference $V_1 -V_0$ or $b$ may play the role 
of a bias voltage. 

\begin{figure}
 \centerline{\includegraphics[width=0.6\textwidth]{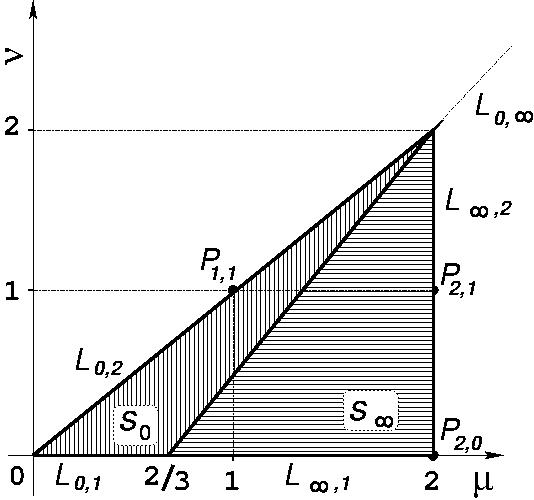}}
%\vspace{0.1pt}
\caption{ Regions of asymptotic representations $z_0, z_1 \to 0$ ($S_0$)
and $z_0, z_1 \to \pm \, \infty$ ($S_\infty$) with separating line $L_{0, \infty}$. 
Three balls indicate  characteristic points $P_{1,1} := \{ \mu= \nu=1\} 
\in S_{0}\,$, $P_{2,0} := \{ \mu =2, \, \nu =0\} \in S_\infty$ and 
$P_{2,1} := \{ \mu=2,\, \nu=1\} \in S_\infty$.}
\label{fig2}
\end{figure}

Due to Equations (\ref{33}),  we have the asymptotics  
 $z_0, z_1 \sim \varepsilon^{2(1 + \nu)/3 - \mu }$. Consequently,
  the line $L_{0,\infty}:= \{ 0< \mu \le 2, \, \nu = 3\mu/2 -1 \}$  separates 
the asymptotic representations $z_0, z_1 \to 0 $ and $z_0, z_1 \to \pm \, 
\infty$ on the ($\mu, \nu$)-plane as illustrated by the diagram depicted
in {\bf Figure \ref{fig2}}. 
%-----------------------------fig2--------------------------------------
%-----------------------------fig2--------------------------------------
Here, we have the  two triangle sets:
\begin{eqnarray}
S_0 &:= & \{  0 < \mu < 2, \,
\max\{ 0,\, 3\mu/2 - 1\} < \nu \le \mu \} \cup \{ 0< \mu < 2/3, \, \nu = 0\},
 \nonumber \\
S_\infty &:= &  \{  2/3 < \mu \le 2, \, 0 \le \nu < 3\mu/2 -1 \},  
 \label{58}
 \end{eqnarray}
  where the asymptotic representations $z_0, z_1 \to 0$ and $z_0, z_1 \to \pm \,
  \infty$ take place, respectively. The corresponding 
  angles are formed by the boundary 
  lines: $S_0$ by $L_{0,1} := \{ 0< \mu < 2/3,\, \nu =0 \}$, 
  $L_{0,2} := \{ 0 < \mu < 2,\, \nu = \mu \}$ and $S_\infty$ by 
  $L_{\infty , 1} := \{ 2/3 < \mu \le 2,\, \nu =0 \}$, 
  $L_{\infty , 2} := \{\mu =2,\, 0 < \nu < 2 \}$.

\subsection{Point interactions realized in the limit as $z_0, z_1 \to 0$ }
 
 Let us consider the boundary lines $L_{0,1}$ and $L_{0,2}$ 
  of the angle $S_0$. On the line $L_{0,1}$,   we find
 that $z_0, z_1 \sim \varepsilon^{2/3 -\mu}$, so that the $z_0, z_1 \to 0$ limit
 takes place on the interval $0 < \mu < 2/3$. Next, we have 
 $k_0^2, k_1^2 \sim \varepsilon^{- \mu}$ and according to Equations (\ref{41}),
 $\lambda_{12} \to 0$ and $\lambda_{21} \sim \varepsilon^{1- \mu} \to 0$,
 so that the $\Lambda$-matrix becomes the identity ($\Lambda = I$)
  because $\mu < 1$.
  
 Similarly, on the  line $L_{0,2}\,$,   where
$z_0, z_1 \sim \varepsilon^{(2 -\mu)/3}$, from  Equations (\ref{39}) and (\ref{41})
we get the  asymptotics
\begin{eqnarray}
\lambda_{11} & \to & 1 - \, c_{1} \varepsilon^{2- \mu},
~~ \lambda_{12} \to \varepsilon d, \nonumber \\ 
\lambda_{21} &\to &  (a + b/2)d \,  \varepsilon^{1- \mu}, ~~
\lambda_{22}  \to  1 - c_{2}\, \varepsilon^{2- \mu}
\label{59}
\end{eqnarray}
with
\begin{equation}
c_{1}:= (a^2/2)(a+ b)(d/b)^2,~~c_{2}:= (a/2)(a+ b)^2(d/b)^2 .
\label{60}
\end{equation}

Therefore, on the interval $0< \mu < 1$ the transmission matrix is the identity $I$,
while on the interval $1 < \mu < 2$ the transmission matrix does not exist.
In this case the point interaction acts as a fully reflecting wall 
(the boundary conditions 
for this point interaction are of the Dirichlet type). 
The value $\mu =1$ describes the intermediate situation with a partial transmission
through the system, namely
the $\delta$-interaction with bias $b$, which separates both these regimes.   
The limit transmission matrix (as $\varepsilon \to 0$)
corresponds to the $\delta$-interaction described by the connection matrix
(\ref{1u}) with the strength constant 
\begin{equation}
\alpha =  (a + b/2)d.
\label{61}
\end{equation}
 This result also includes the  constant
case when $V_0 = V_1$, i.e., $b =0$.  This approximation is appropriate
for modeling the $\delta$-potential. Note that  similar analysis 
can be done for $\mu$ and $\nu$ belonging to the interior of $S_0$.
In this case in the above equations we have to set $b =0$.

Using  the second formula (\ref{27}), one can compute the transmission amplitude
for  this $\delta$-interaction. We get
\begin{equation}
{\cal T }= {4k\,k_R  \over ( k + k_R)^2 + \alpha^2 },
\label{62}
\end{equation}
where $\alpha$ is given by (\ref{61}). 
In the unbiased case ($b=0,~k_R= k$) this formula reduces to 
${\cal T} = \left[1+ (\alpha/2k)^2\right]^{-1}$ with $\alpha =ad$,
 the well known expression 
for  the  constant potential. 
Equation (\ref{62}) has been obtained for any  $a \in \R$.
 However, for negative values of $a $, i.e., for a $\delta$-like well, it does not
 describe the oscillating behavior with respect to the constant $\alpha$ 
 that takes place  under tunneling across a  well with finite thickness $l$.

 \subsection{Point interactions realized in the limit as 
 $z_0, z_1 \to \pm \, \infty$ }
 
Consider now the  characteristic point $P_{2,1} \in S_\infty$ setting in
Equations (\ref{52}) $\mu =2$ and $\nu =1$. 
   Here $k_0^2 - k_1^2 = V_1 -V_0 = b \, \varepsilon^{- 1}$ and $k_0, k_1, 
k_{1,0} \to \sqrt{-\, a}\, \varepsilon^{-1}$,  so that 
 the asymptotic representation  of Equations (\ref{52}) in the limit as 
 $\varepsilon \to 0$ becomes  
\begin{eqnarray}
\lambda_{11} & \to & \cos(\kappa  d)- \varepsilon \, g
\sin(\kappa d), \nonumber \\
\lambda_{12} & \to &  \varepsilon \,\kappa^{-1}\sin(\kappa d), \nonumber \\
\lambda_{21} &\to & -\, \varepsilon^{-1} \kappa \sin(\kappa d) + 
{\cal O} (\varepsilon), \nonumber \\
\lambda_{22} &\to & \cos(\kappa d)+ 
\varepsilon \,g\sin(\kappa d),
\label{63}
\end{eqnarray}
where 
\begin{equation}
\kappa : = \sqrt{-\, a} , ~~~g := \kappa^{-3}(b/4d).
\label{64}
\end{equation}

As follows from these asymptotic expressions derived at the point $P_{2,1}$, 
in the limit as $\varepsilon \to 0$,  the transmission through a barrier is
zero, while across a well ($ a < 0$) it 
appears to be resonant. The resonance set consists of the roots of 
the equation $\sin(\kappa d) =0$. At fixed $d >0$, these roots form
the countable set $\Sigma = \cup_{n=0}^\infty \sigma_n$ formed from 
 the points $\sigma_n := -\, (n \pi /d)^2$. On this resonance set, the
 discrete-valued matrix is  $\Lambda_n : =\Lambda|_\Sigma = (- 1)^n I$. 
Beyond these resonance values, the $\delta$-like well is opaque
and, instead of the identity matrix $I$, the two-sided boundary conditions 
for the wave function are of the Dirichlet type ($\psi(\pm \, 0)= 0$).

\section{  Multi-layered heterostructures with bias }

Now we are ready to apply the expressions obtained above for a single layer
to the total structure consisting of an arbitrary number $N$ of layers
replacing $\mu \to  \mu_i\,, ~\nu \to \nu_i\,, ~b \to b_i\,, ~d \to d_i\,.$
Taking for account that the left boundary value for the potential of
 the $i$th layer $a_i$ is shifted because of the biases  $b_1\,, \ldots b_{i-1}$
 in the left-hand layers, we need to use the following replacement rule:
\begin{equation}
a \to a_i + \sum_{j=1}^{i-1} b_j\,, ~~i=\overline{1,N},
\label{65a}
\end{equation} 
 where the sum vanishes if $i=1$. Then Equations (\ref{57}) are transformed to
 \begin{equation}
  V_0 \to  V_{i,i-1}= \left( a_i + \sum_{j=1}^{i-1}b_{j} \right) \! 
\varepsilon^{- \mu_i},~~
V_1 \to V_{i,i} = V_{i, i-1} + b_i \,\varepsilon^{-\, \nu_i}.
 \label{65}
 \end{equation}
 Next, all the other expressions derived above
 should be   rewritten  for the  $i$th layer using the following 
replacement rules: 
\begin{eqnarray} 
 z_0 &\to & z_{i,i-1} =\left( { d_i \over b_{i}}\right)^{\!\!2/3}
\left[ \left( a_i +\sum_{j=1}^{i-1}b_{j}\right)
\! \varepsilon^{- \mu_i} -E \right]\!\varepsilon^{2(1+\nu_i)/3},\nonumber \\
 z_1 & \to & z_{i,i} = \left({ d_i \over b_{i}}\right)^{\!\!2/3}
\left[ \left( a_i +\sum_{j=1}^{i-1}b_{j}\right)
\!\varepsilon^{- \mu_i} + b_{i}\,\varepsilon^{- \nu_i} -E \right]\!
\varepsilon^{2(1+\nu_i)/3}, \nonumber \\
 \alpha &\to &  \alpha_i =  
 \left(a_{i} +\sum_{j=1}^{i-1}b_j + b_i/2 \right)\!d_i\,, ~~
\sigma \to \sigma_i= \left( b_i \over d_i \right)^{\!\!\!1/3}\!\!
\varepsilon^{-(1+\nu_i)/3},\nonumber \\
\kappa &\to & \kappa_i = \sqrt{- \left( a_i + \sum_{j=1}^{i-1}b_j\right)},~~
g \to g_i =  {b_i \over 4 \kappa_i^{3} d_i} \,,
\nonumber \\
c_{1} &\to & c_{i,1} = \frac12 \left(  a_i + \sum_{j=1}^{i-1}b_j\right)^{\!\!\!2}
\left(  a_i + \sum_{j=1}^{i}b_j\right)\left({d_i \over b_i}\right)^{\!\!2}, 
\nonumber \\
c_{2} &\to & c_{i,2} = \frac12 \left(  a_i + \sum_{j=1}^{i-1}b_j\right)
\left(  a_i + \sum_{j=1}^{i}b_j\right)^{\!\!\!2} \left({d_i \over b_i}\right)^{\!\!2}.
\label{66}
\end{eqnarray}

In the following  we will consider some  particular examples of 
multi-layered structures with $N=2,3$. It will be shown that in some cases 
two- and three-lateral 
quantum devices  can be approximated by one-point interactions. 

\subsection{Two-layered structures}

\begin{figure}
\centerline{\includegraphics[width=0.5\textwidth]{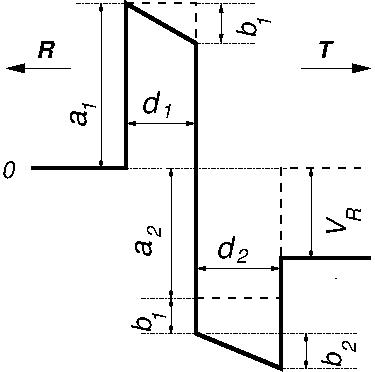}}
%\vspace{0.1pt}
\caption{ Schematics of tilted (solid line) and piecewise constant (dashed line)
  barrier-well potential, where  notations correspond to Equations (\ref{65a}) and
  (\ref{65}) for $N=2$ and  $\varepsilon =1$. Potential values at layer edges 
 are $V_{1,0}= a_1$, $V_{1,1} = a_1 +b_1$ (barrier, $a_1 >0$) and 
 $V_{2,1} = a_2 +b_1$, $V_{2,2} = a_2 + b_1 + b_2$ (well, $a_2 <0$).
Polarity is shown  positive (left-to-right electron flow, $b_1\, , b_2 <0$).  
 Dashed lines show unbiased potential ($b_1 =b_2 =0$).   }
\label{fig3}
\end{figure}

Consider now the structure consisting of  two layers ($N=2$).
The piecewise linear potential of a barrier-well form is shown in 
{\bf Figure \ref{fig3}}.
%-----------------------------fig3--------------------------------------
%-------------------------fig3------------------------------------ 
For an arbitrary two-layered structure,  the limit transmission
matrix is the product $\Lambda = \Lambda_2 \Lambda_1$, where the matrices 
$\Lambda_i$'s can be constructed from the asymptotic approximations (\ref{59})
and (\ref{63}) by applying the replacement rules (\ref{65a})-(\ref{66}). 
Applying these rules in
Equations (\ref{59}), (\ref{60}) and (\ref{63}), (\ref{64}) to the matrices
$\Lambda_1$ and $\Lambda_2$, 
below we compute their product for two different situations.  
Note that due to the presence of the factor $\varepsilon^{-1}$
in the expression $\lambda_{21}$ [see Equations (\ref{63})], the terms of 
order ${\cal O}(\varepsilon)$ must be kept in the product
$\Lambda_2\Lambda_1$  because $\lim_{\varepsilon \to 0}\Lambda_2 \cdot
\lim_{\varepsilon \to 0}\Lambda_1 \neq \lim_{\varepsilon \to 0}
(\Lambda_2\Lambda_1)$.

 {\em Point interactions of a $\delta'$-type:}
  Consider  the zero-thickness limit determined by the powers
  $\mu_1 = \mu_2 =2$ and $\nu_1 = \nu_2 =1$. Then, 
 the product $\Lambda= \Lambda_2\Lambda_1$
 yields the following asymptotic representation
  of the $\Lambda$-matrix elements for the total double-layer system:
    \begin{eqnarray}
\lambda_{11} & \to & \cos(\kappa_1 d_1)  \cos(\kappa_2 d_2) -  (\kappa_1/\kappa_2)
\sin(\kappa_1 d_1) \sin(\kappa_2 d_2) ,\nonumber \\
 \lambda_{12} & \to & 0,  \nonumber \\
 \lambda_{21} & \to & \alpha -\, \varepsilon^{-1}
 \left[\kappa_1  \sin(\kappa_1 d_1) \cos(\kappa_2 d_2)   +
  \kappa_2  \cos(\kappa_1 d_1)  \sin(\kappa_2 d_2)  \right] ,  \nonumber \\
 \lambda_{22} & = &  \cos(\kappa_1 d_1) \cos(\kappa_2 d_2) 
 - \,(\kappa_2/\kappa_1) \sin(\kappa_1 d_1)  \sin(\kappa_2 d_2) ,
\label{67}
\end{eqnarray}
where the $\varepsilon \to 0$ limit has been performed and 
\begin{equation}
\alpha = (\kappa_2 g_1 - \kappa_1 g_2)
\sin(\kappa_1 d_1) \sin(\kappa_2 d_2).
\label{68}
\end{equation}
The second term in  the element $\lambda_{21}$ diverges as $\varepsilon \to 0$
and it vanishes if the equation 
\begin{equation}
\kappa_1 \tan(\kappa_1 d_1)  +  \kappa_2 \tan(\kappa_2 d_2)  = 0 
\label{69}
\end{equation}
takes place.
Using this equation in the elements $\lambda_{11}$ and  $\lambda_{22}$
[see Equations (\ref{67})],  we find  the total transmission matrix
 \begin{equation}
\Lambda = \left(\begin{array}{ll}
\cos(\kappa_1 d_1) /\cos(\kappa_2 d_2) ~~~~~~~~~~~ 0 \\ \alpha
 ~~~~~~~~~~~~ \cos(\kappa_2 d_2) /\cos(\kappa_1 d_1)
\end{array} \right) .
\label{70}
\end{equation}

Equation (\ref{69}) admits a countable set of solutions if at least  one of
the layer potential  has a well profile. In particular, if 
 $a_1 >0$ (barrier) and $a_2 + b_1 < 0$ (well),  Equation (\ref{69}) reduces to
\begin{equation}
\sqrt{a_1} \tanh(\sqrt{a_1}\,d_1)= \sqrt{|a_2 + b_1|} 
\tan(\sqrt{|a_2 + b_1|} \, d_2).
\label{71}
\end{equation} 
 It is reasonable to assume that $-\, b_1 < a_1$ (otherwise 
 the right-edge barrier potential  becomes negative), so that
on the interval $(-\, a_1, 0) $, under appropriate values of the layer parameters,
only a finite set of discrete (resonance) values of $b_1$ can be found. 
 According to the classification of point interactions given in \cite{bn},
 the  interactions described by the connection matrix with diagonal 
 elements $\lambda_{11},\, \lambda_{22} \neq 1$ 
 may be referred to as a family of  (resonant) $\delta'$-potentials,
 despite the distribution $\delta'(x)$ in general does not exist.
Similarly to the single $\delta$-well potential, beyond the resonance set,
the two-sided boundary conditions are of the Dirichlet type: $\psi(\pm 0)=0$.

On the resonance set $\Sigma = \cup_n \sigma_n$, the explicit expressions 
for the $\Lambda$-matrix (\ref{70}) and the element (\ref{68}) become
 \begin{equation}
 \Lambda |_{\Sigma} = \left(\begin{array}{ll}  \theta_n ~~~~ 0 \\ \alpha_n
 ~~~ \theta^{-1}_n \end{array} \right) ,
\label{71a}
\end{equation}
where
\begin{eqnarray}
\theta_n &=& { \cosh(\sqrt{a_1}\,d_1) \over  \cos(\sqrt{|a_2 +b_{1,n}|}\, d_2)}
\neq \pm \, 1,
\nonumber \\
\alpha_n & =& \frac14 \left[ {\sqrt{a_1}\, b_2 \over  |a_2 + b_{1,n}|^{3/2}d_2} -
{\sqrt{|a_2 +b_{1,n}|}\, b_{1,n} \over  a_1^{3/2}d_1 }\right] \nonumber \\
 && \times \sinh(\sqrt{a_1}\, d_1) \sin(\sqrt{|a_2 + b_{1,n}|} \,d_2).
\label{71b}
\end{eqnarray}
The transmission amplitude on the resonance set $\Sigma $ is 
\begin{equation}
{\cal T}_n = {4k\,k_{R,n} \over (k \theta_n^{-1} + k_{R,n}\theta_n)^2 +\alpha_n^2} ,
\label{71c}
\end{equation}
where $k_{R,n} = \sqrt{k^2 - b_{1,n}}\,$.

{\em Resonant transmission through a $\delta$-barrier:} Let us 
consider now the two-layered structure in which the potential of one of the layers 
in the squeezed limit has a $\delta$-like form. We specify this situation
by the power parameters $\mu_1 =\nu_1 =1$ (point  $P_{1,1} 
\in S_0$) for the barrier, and $\mu_2 =2$ and $\nu_2 =1$ (point $P_{2,1}
 \in S_\infty$) for the well.   Even in the 
unbiased case this potential  has no distributional limit, however the transmission matrix does exist. Applying the 
replacement rules (\ref{65a})-(\ref{66}) in the asymptotics  (\ref{59})
with $\mu =1$  yielding the $\Lambda_1$-matrix, and in the 
representation (\ref{63}) creating the $\Lambda_2$-matrix,  
 we obtain the $\varepsilon \to 0$ limit for  the elements of the total matrix
 $\Lambda = \Lambda_2 \Lambda_1$  in the form
\begin{eqnarray}
\lambda_{11}& \to &   \cos(\kappa_2 d_2), \nonumber \\
\lambda_{12} &\to & 0, \nonumber \\
\lambda_{21} & \to &   \alpha_1 \cos(\kappa_2 d_2) 
+ c_{1,1} \kappa_2 \sin(\kappa_2 d_2) - \varepsilon^{-1}
\kappa_2 \sin(\kappa_2 d_2), \nonumber \\
\lambda_{22}& \to &   \cos(\kappa_2 d_2) -
\kappa_2 d_1 \sin(\kappa_2 d_2).
\label{72}
\end{eqnarray}
 While the first and the second terms in $\lambda_{21}$ are finite, the 
 third  one diverges as $\varepsilon \to 0$. However, it vanishes 
 at the values satisfying the equation $\sin(\kappa_2 d_2) =0$, i.e., for
 \begin{equation}
 a_2 + b_{1,n} = -\, (n\pi/d_2)^2,
 \label{73}
 \end{equation}
 where the integer $n =n_0, n_0 +1, \ldots $ with some $n_0$.  These values
 form the countable resonance set $\Sigma$ on which the transmission matrix $\Lambda$
 corresponds to the $\delta$-interaction, whereas beyond this set the 
 interaction acts as a fully reflecting wall. The limit
 transmission matrix is
 \begin{equation}
 \Lambda |_{\Sigma} = (- 1)^n \left( \begin{array}{ll} 1 ~~~~\, 0 \\ 
 \alpha_n  ~~~1 \end{array} \right) , 
\label{73a}
\end{equation} 
where $\alpha_n = \alpha_{1,n} = (a_1 +b_{1,n})/2)d_1\,.$ Note that the
   effect of the resonant 
 transmission through a $\delta$-barrier keeps to be valid
  in the unbiased case when $b_1 =b_2 =0$.
 
 Thus,  we have realized
 the {\em resonant} $\delta$-interaction, due to the presence of an adjacent
 well with depth $a_2 <0$. In the case when the system parameters
 $a_1, a_2, d_1, d_2$ are supposed to be fixed, the biased potential 
 $b_1$ may be considered as a  tunable parameter. 
   The transmission is resonant on the set given by (\ref{73}).
 The  potential at the right edge of the first layer 
  keeps to be positive for all values of $b_1$ satisfying the inequality
$-\, b_1 < a_1$. Therefore this is  a constraint that limits the
resonance set to a finite number of resonances. 

\begin{figure}
\centerline{\includegraphics[width=0.8\textwidth]{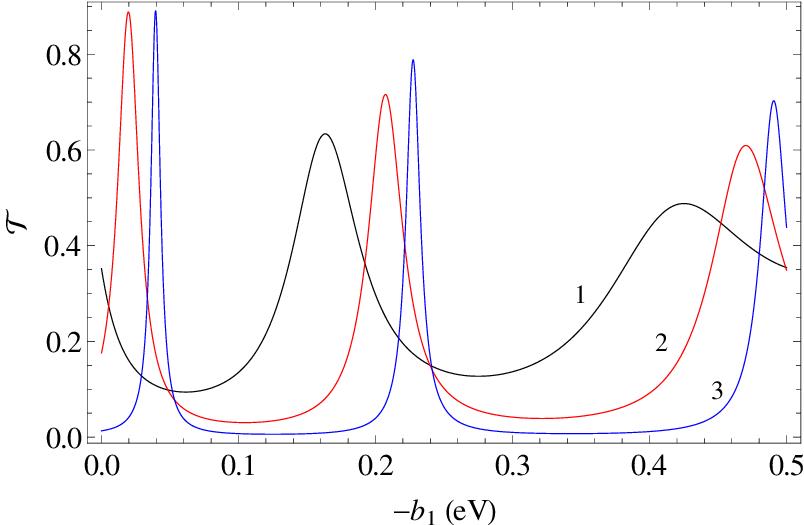}}
%\vspace{0.1pt}
\caption{ Transmission amplitude ${\cal T}$ as a function of 
bias $-\,b_1$ plotted 
for parameter values: $E = 0.1$ eV, $a_1 = 0.5$ eV, 
$a_2 =-\, 0.1$ eV, $d_1 =2$ nm, 
$d_2 = 10$ nm.  Computations have been carried out with  powers
 $\mu_1 = \nu_1=1$ (point $P_{1,1}$) and $\mu_2= 2$, $\nu_2 =1$ (point $P_{2,1}$).
  Squeezing scenario is displayed for  $\varepsilon =0.5$ (curve 1, black),
  $0.25$ (curve 2, red) and $0.1$ (curve 3, blue). 
  Location of all three peaks
 converges  to set $\{-\, b_{1,n}\}$ defined  by 
  Equation (\ref{73}) with $n =2,\,3,\,4$. 
 }
\label{fig4}
\end{figure}

The existence of the resonant tunneling through a $\delta$-like barrier 
can be supported numerically calculating the transmission amplitude ${\cal T}$
according to Equations (\ref{27}) and (\ref{28}), where the matrix elements 
are given by Equations (\ref{35}). For different values of the squeezing 
parameter $\varepsilon$, the result of these calculations is illustrated
by {\bf Figure \ref{fig4}}.
%-----------------------------fig4--------------------------------------
%-------------------------fig4------------------------------------ 

\subsection{Modeling of point transistors}

\begin{figure}
\centerline{\includegraphics[width=0.8\textwidth]{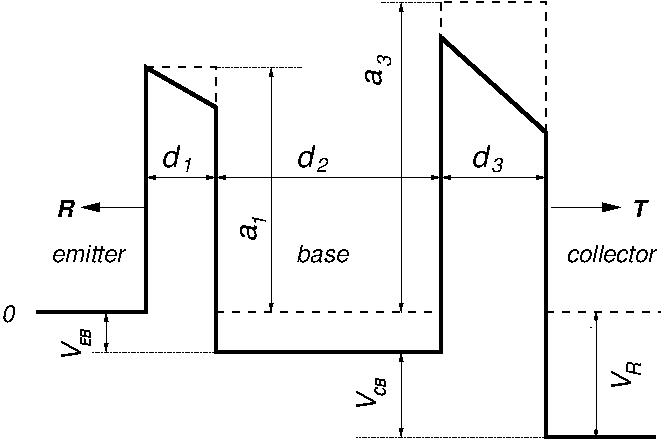}}
%\vspace{0.1pt}
\caption{ Schematics of typical transistor, where  notations correspond to 
Equations (\ref{65a}) and  (\ref{65}) for $N=3$ and  $\varepsilon =1$
with replacement: 
$b_1 \to -\, V_{EB}$ (emitter-to-base voltage) and  
$b_3 \to -\, V_{CB}$ (collector-to-base voltage). 
Potential values at layer edges 
 are $V_{1,0} = a_1$, $V_{1,1} = a_1 - V_{EB}$ ($a_1 >0$), 
 $V_{2,1} = V_{2,2}= -\, V_{EB}$ and  $V_{3,2} = a_3 - V_{EB}$,
 $V_{3,3} = a_3 + V_{R}$ ($a_3 >0$).
Polarity is shown to be positive (left-to-right electron flow, $V_{EB}\, ,
V_{CB} > 0$).
  }
\label{fig5}
\end{figure}

It is of interest to give an interpretation 
for a semiconductor transistor in the limit
as its dimensions are extremely tiny. This is a three-terminal device \cite{gw}
described by  a tilted double-barrier potential profile as illustrated  
by {\bf Figure \ref{fig5}}. Here the potential between the barriers is constant
 depending on the emitter-to-base voltage $V_{EB}$ as a parameter tuned externally. 
 The other external  parameter $V_{CB}$ is the collector-to-base voltage being
  fixed.  For the description of this device by a one-point interaction model,
  we assume that in the zero-thickness 
limit both the barriers as well as the distance between them
tend to the point $x=0$. 
%-----------------------------fig5--------------------------------------
%-------------------------fig5------------------------------------ 

Similarly to the double-layer structure [see the general formula (\ref{18})], 
the transmission matrix for the total system is
the product $\Lambda= \Lambda_3\Lambda_2\Lambda_1$, where
the matrices $\Lambda_1$ and $\Lambda_3$ correspond to the barriers
and the $\Lambda_2$-matrix to the space between the barriers. 
Setting $b_1 \equiv - \,V_{EB}$, $b_2 =0$ and $b_3 \equiv -\, V_{CB}$, 
 according to (\ref{65a}), we  replace:
  $a \to a_1$  for $\Lambda_1$, $a \to -\, V_{EB}$ ($a_2 =0$) for $\Lambda_2$
and $a \to a_3 -V_{EB}$ for $\Lambda_3$. In the case of positive polarity, 
as shown in the figure,   both the voltages are non-negative parameters.  
Applying next the replacement rules  (\ref{66}) in the terms (\ref{60}),
(\ref{61}) and (\ref{64}), we write the following explicit expressions 
for the matrices $\Lambda_1$ and $\Lambda_3$: 
\begin{eqnarray}
\kappa_1 &=& \sqrt{-\, a_1}\,,~~\kappa_2 = \sqrt{V_{EB}}\,,~~\kappa_3 = 
\sqrt{V_{EB} - a_3 }\,, \nonumber \\
\alpha_1 &=& (a_1 - V_{EB}/2) d_1\,, ~~\alpha_3 = (a_3 - V_{EB} - V_{CB}/2)d_3\,,
\nonumber \\
c_{1,1 } & =&   (a_1^2/2)(a_1 - V_{EB})(d_1/V_{EB})^2, \nonumber \\
 c_{1,2} & = &  (a_1/2)(a_1 - V_{EB})^2(d_1/V_{EB})^2  , \nonumber \\
c_{3,1 }& = &  [(a_3 - V_{EB})^2/2](a_3 - V_{EB} - V_{CB})(d_3/V_{CB})^2,
\nonumber \\
 c_{3,2} & =&  [(a_3 - V_{EB})/2](a_3 - V_{EB} - V_{CB})^2(d_3/V_{CB})^2 ,
 \nonumber \\
 g_1 & =& -\, \kappa_1^{- 3}(V_{EB}/4d_1),~~ g_3 = -\, \kappa_3^{- 3}(V_{CB}/4d_3).
\label{74}
\end{eqnarray}
The $\Lambda_2$-matrix  is defined by
 (\ref{38}) for $i=2$, where $k_2 = \kappa_2 \varepsilon^{-1} $
and $l_2 =\varepsilon d_2$.

Below we examine the  following two zero-thickness limits: (i)
$\mu_1 = \mu_3 =\nu_1 =\nu_3 =1 $ (points $P_{1,1}$) and 
(ii)  $\mu_1 = \mu_3 = 2$, $\nu_1 =\nu_3 =1 $ (points $P_{2,1}$).

{\em (i) $\delta$-potential model:}
The matrix multiplication yields the asymptotic representation in the limit
 as $\varepsilon \to 0$:  
\begin{eqnarray}
\lambda_{11} & \to &  \cos(\kappa_2 d_2)  -  
 \kappa_2 d_3 \sin(\kappa_2 d_2),  \nonumber \\
 \lambda_{12} &\to & 0, \nonumber \\
\lambda_{21} & \to & 
 (\alpha_1 + \alpha_3 )\cos(\kappa_2 d_2)
 + (c_{1,1} + c_{3,2}) \kappa_2  \sin(\kappa_2 d_2)
 - \varepsilon^{-1}  \kappa_2 \sin(\kappa_2 d_2), \nonumber \\
\lambda_{22} & \to & \cos(\kappa_2 d_2) -  \kappa_2 d_1  \sin(\kappa_2d_2).
\label{75}
\end{eqnarray}
Here, the element $\lambda_{21}$ diverges as $\varepsilon \to 0$ and it
will be finite if $\sin(\kappa_2 d_2) =0$, resulting in  the resonance set
\begin{equation}
V_{EB,n} = (n \pi/d_2)^2,~~n= \overline{1, n_0}\, ,
\label{76}
\end{equation}
where the integer $n_0$ depends on the interval of admissible values of
the bias potential $V_{EB}$. This interval is determined by the requirement 
that the barrier potential values $V_{1,1}$ and $V_{3,3}$ must be positive,
leading to the inequalities $0 < V_{EB} < a_1$ and $0 < V_{EB} + V_{CB} < a_3$. 
Therefore the potential $V_{EB}$ is allowed to tune within the interval 
$0 < V_{EB} < \min\{ a_1,\, a_3 - V_{CB}\}$.

Thus,  the limit transmission matrix is of the form (\ref{73a}) with
\begin{equation}
\alpha_n = \alpha_{1,n} +\alpha_{3,n} = (a_1 - V_{EB,n}/2) d_1 + 
(a_3 - V_{EB,n}- V_{CB}/2) d_3 >0.
\label{77}
\end{equation}
realizing the $\delta$-potential defined on the resonance set described by
Equation (\ref{76}).

According to the general expressions (\ref{27}) and (\ref{28}), 
the transmission amplitude, being non-zero on this resonance set, is given by 
the formula (\ref{71c}),
where $\theta_n =1$ and $k_{R,n}= \sqrt{k^2 + V_{EB,n} + V_{CB}}\,$.
%-----------------------------fig6--------------------------------------
\begin{figure}
\centerline{\includegraphics[width=0.8\textwidth]{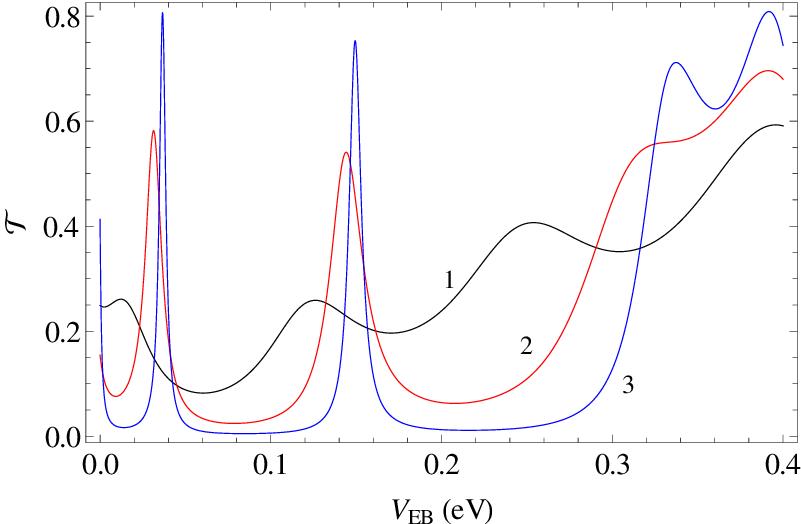}}
%\vspace{0.1pt}
\caption{ Transmission amplitude ${\cal T}$ as a function of 
emitter-to-base voltage $V_{EB}$ for parameter values: $E= 0.1$ eV,
 $a_1 = a_3 = 0.5$ eV, $a_2 =0$, $V_{CB} = 0.2$ eV,
 $d_1 =d_3 =2$ nm, $d_2 = 10$ nm. Computations have been carried out 
 with  powers  $\mu_1 =\nu_1 = \mu_3 = \nu_3 =1$ (points $P_{1,1}$)
 and $\mu_2=2$, $\nu_2 =0$. Squeezing scenario is displayed for
  $\varepsilon =0.5$ (curve 1, black), $0.25$ (curve 2, red) and 
  $0.1$ (curve 3, blue). Location of all three peaks approaches 
 set $\{ V_{EB,n} \}$ given by Equation (\ref{76}) with $n =1,\,2,\,3$. 
  }
\label{fig6}
\end{figure}
%-------------------------fig6------------------------------------ 
The transmission amplitude ${\cal T}$ displayed in {\bf Figure \ref{fig6}} 
illustrates the convergence of the location of the peaks to the roots of 
Equation (\ref{76}).  

{\em (ii) $\delta'$-potential model:}
The three-lateral device can also  be approximated by a $\delta'$-interaction
with a bias if we choose for the zero-thickness 
limit the powers $\mu_1 = \mu_2 = \mu_3 = 2$
and $\nu_1 = \nu_3 =1$.
The multiplication of the matrices yields
\begin{eqnarray}
\lambda_{11} & \to & \cos(\kappa_1 d_1) \cos(\kappa_2 d_2) \cos(\kappa_3d_3) 
-(\kappa_1/\kappa_2)\sin(\kappa_1 d_1)  \sin(\kappa_2 d_2)  \cos(\kappa_3d_3)  \nonumber \\
&& -\, (\kappa_1/\kappa_3)\sin(\kappa_1 d_1) \cos(\kappa_2 d_2)
 \sin(\kappa_3d_3)   \nonumber \\
&& -(\kappa_2/\kappa_3)\cos(\kappa_1 d_1)  \sin(\kappa_2 d_2) 
 \sin(\kappa_3d_3) , 
\nonumber \\
\lambda_{12} &\to & 0 , 
\nonumber \\
\lambda_{21} &\to & 
   \kappa_2 [ g_{1} \sin(\kappa_1 d_1)  \cos(\kappa_3d_3) 
 - g_{3} \cos(\kappa_1 d_1)  \sin(\kappa_3d_3)  ]
 \sin(\kappa_2 d_2)  \nonumber \\
 && + \, (\kappa_3 g_{1} - \kappa_1 g_{3}   ) \sin(\kappa_1 d_1) 
  \cos(\kappa_2 d_2)  \sin(\kappa_3d_3)  
\nonumber \\
&& -\,\varepsilon^{-1} [ \kappa_1 \sin(\kappa_1 d_1) 
 \cos(\kappa_2 d_2)  \cos(\kappa_3d_3)  \nonumber \\
 && + \, \kappa_2 \cos(\kappa_1 d_1)
  \sin(\kappa_2 d_2)  \cos(\kappa_3d_3)   
 +  \kappa_3 \cos(\kappa_1 d_1)  \cos(\kappa_2 d_2)  \sin(\kappa_3d_3) \nonumber \\
&& -\, (\kappa_1\kappa_3/\kappa_2)\sin(\kappa_1 d_1) \sin(\kappa_2 d_2) 
 \sin(\kappa_3d_3) ], \nonumber \\
\lambda_{22} &\to & \cos(\kappa_1 d_1) \cos(\kappa_2 d_2)  \cos(\kappa_3d_3)   
-(\kappa_2/\kappa_1)\sin(\kappa_1 d_1)  \sin(\kappa_2 d_2)  \cos(\kappa_3d_3)   \nonumber \\
& & -\, (\kappa_3/\kappa_1)\sin(\kappa_1 d_1)  \cos(\kappa_2 d_2) 
 \sin(\kappa_3d_3) \nonumber \\
&& -\,(\kappa_3/\kappa_2)\cos(\kappa_1 d_1)  \sin(\kappa_2 d_2) \sin(\kappa_3d_3),
\label{79}
\end{eqnarray}
where the notations for  $\kappa_1\,,\kappa_2\,,\kappa_3$ and 
 $g_{1}\,, g_{3}$ can be found in Equations (\ref{74}). 
The arguments of the trigonometric functions 
are finite and  the element $\lambda_{21}$ diverges 
as $\varepsilon \to 0$ because of the presence of the factor $\varepsilon^{-1}$. 
 Therefore the only opportunity to 
define properly a point interaction is a full cancellation of all the terms
at this factor, so that $\lambda_{21}$ becomes  finite. As a result,
this cancellation yields the following equation:
\begin{equation}
 { \kappa_1 \kappa_3 \over \kappa_2} \prod_{i =1}^3\tan(\kappa_i d_i)
  = \sum_{i=1}^3  \kappa_i \tan(\kappa_i d_i) .
  \label{80}
  \end{equation}
  Using the resonance equation (\ref{80}), we derive that the 
   pair $\{\lambda_{11}, \lambda_{22}\}$
admits the following  sixteen representations: 
\begin{equation}
\{\lambda_{11}, \lambda_{22}\} =
\{I_1, I_2, J_1^{-1}, J_2^{-1}\} \times \{I_1^{-1}, I_2^{-1}, J_1, J_2 \}, 
\label{81}
\end{equation}
where  
\begin{eqnarray}
I_1 & = & {\cos(\kappa_1 d_1) \cos(\kappa_2 d_2)  - (\kappa_1/\kappa_2) 
\sin(\kappa_1 d_1)   \sin(\kappa_2 d_2)   \over \cos(\kappa_3 d_3)   } \,,\nonumber \\
I_2 &=& -\, { \kappa_1 \sin(\kappa_1 d_1)  \cos(\kappa_2 d_2)   + 
 \kappa_2 \cos(\kappa_1 d_1)    \sin(\kappa_2 d_2) 
   \over \kappa_3 \sin(\kappa_3 d_3)   } \, ,\nonumber \\
J_1 &=& {\cos(\kappa_2 d_2)   \cos(\kappa_3 d_3)   - (\kappa_3/\kappa_2)  
\sin(\kappa_2 d_2)   \sin(\kappa_3 d_3)  \over \cos(\kappa_1 d_1)  }\,, 
\nonumber \\
J_2 &=& -\, { \kappa_2 \sin(\kappa_2 d_2) \cos(\kappa_3 d_3)    + 
\kappa_3 \cos(\kappa_2 d_2) \sin(\kappa_3 d_3)
  \over \kappa_1 \sin(\kappa_1 d_1) }\,.
\label{82}
\end{eqnarray}
These representations follow from the equations 
 $I_1 = I_2$, $J_1 =J_2$ and $I_1 J_1 =1$, which can be checked using
  the condition (\ref{80}). As a result, we have 
  $\det\Lambda=\lambda_{11}\lambda_{22}=1 $ if  Equation (\ref{80}) is fulfilled.
 
Equation (\ref{80}) can be rewritten in the explicit form as follows
\begin{eqnarray}
 &&  \sqrt{a_1/ V_{EB}}\, \tanh(\sqrt{a_1}\,d_1) + \sqrt{a_3/V_{EB} -1}\,
 \tanh(\sqrt{a_3 - V_{EB}}\,d_3) \nonumber \\
  &&= \left[1 - \sqrt{a_1 / V_{EB}}\, \sqrt{{a_3 / V_{EB}} -1}\,
  \tanh(\sqrt{a_1}\,d_1) \tanh(\sqrt{a_3 - V_{EB}}\,d_3)
 \right] \nonumber \\
 && \times\tan(\sqrt{V_{EB}}\,d_2).
\label{83} 
\end{eqnarray}
This form shows the existence of the roots forming
a resonance set $\Sigma = \{ V_{EB,n} \}$. Inserting next these roots into 
 Equations (\ref{82}), one can get the discrete 
values of the diagonal elements  $\lambda_{11,n}$ 
and $\lambda_{22,n}$ of the matrix set $ \Lambda |_\Sigma $.
 One can write then
$\theta_n := \lambda_{11,n} =\lambda_{22,n}^{-1}= I_{1,n}=I_{2,n}=J_{1,n}^{-1}
=J_{2,n}^{-1}$.
Finally, one can represent the off-diagonal element $\lambda_{21,n}
 = \alpha_n$ as
\begin{eqnarray}
\alpha_n &= & a_1^{-3/2}(V_{EB,n}/4d_1)\sinh(\sqrt{a_1}\, d_1) \nonumber \\
&\times&
[\sqrt{V_{EB,n}}\cosh(\sqrt{a_3 -V_{EB,n}}\,d_3) \sin(\sqrt{V_{EB,n}}\, d_2)
\nonumber \\
&-& \sqrt{a_3 -V_{EB,n}}\sinh(\sqrt{a_3 -V_{EB,n}}\,d_3)  
\cos(\sqrt{V_{EB,n}}\, d_2)] \nonumber \\
&-& (a_3 -V_{EB,n})^{-3/2}(V_{CB}/4d_3)\sinh(\sqrt{a_3 - V_{EB,n}}\, d_3) 
\nonumber \\
&\times& [\sqrt{V_{EB,n}}\cosh(\sqrt{a_1}\,d_1) \sin(\sqrt{V_{EB,n}}\, d_2)
\nonumber \\
&-& \sqrt{a_1}\sinh(\sqrt{a_1}\,d_1)  \cos(\sqrt{V_{EB,n}}\, d_2)].
\label{84}
\end{eqnarray}
Similarly to the double-layer structure with the limit transmission 
matrix (\ref{71a}), we refer  this one-point interaction to as 
the $\delta'$-potential because $\lambda_{11,n}\,,\, \lambda_{22,n} \neq 1$.
The transmission amplitude is given by the same formula (\ref{71c}) in which 
$\theta_n = \lambda_{11,n}$ and 
$\alpha_n = \lambda_{21,n}$ is given by the expression (\ref{84}).

\section{Concluding remarks}

In the present work we addressed the  family of point interactions 
as the zero-thickness limit of  heterostructures composed of several layers. The latter have
 energy diagrams stemming from  tilted linear potentials  that arise as a result of the application of external electric fields. 
 The analysis starts from the solution of the one-dimensional
 stationary  Schr\"{o}dinger
 equation for the structure with finite size using the transfer matrix approach.
Within this approach, we find the transmission matrices for each layer; their 
product quantifies the  penetration amplitude of electrons through the whole system. 
In order to realize point interactions we introduce a squeezing parameter 
$\varepsilon >0$
in the structural parameters  of the system (layer width, potentials at layer edges, etc.)
leading to shrinking the thickness of the system as $\varepsilon \to 0$.
In this limit the potential values at the interfaces of layers must go to infinity 
if we wish to create a point interaction in the squeezed limit. 
At  $\varepsilon= 1$, the structural parameters  correspond to realistic 
values of the device. 

One of interesting features discovered  in the previous publications 
\cite{zz15jpa,c-g,zci,tn,gm,gh,gh1} 
is the appearance of electron  tunneling through one-point barriers that occurs
at some discrete values of system parameters, whereas beyond these values
the system behaves as a fully reflecting wall. The origin of this phenomenon
is  an oscillating behavior of particle transmission. Surprisingly, as 
the system shrinks to a point, the oscillating regular function that describes 
the transmission amplitude, converges pointwise to the function with non-zero
finite  values only at some discrete points in the space of system parameters,
whereas beyond this (resonance) set, the system acts a fully reflecting wall
(see, e.g.,  Figure~1 in \cite{zz15jpa}). In other words, 
 the maxima of the oscillating amplitude correspond in the squeezing limit to
  the set of extremely sharp peaks.   On the other hand, in many devices
  the oscillating behavior of transmitted particles appears as a function of
  tuning some controllable (not system) parameters. For instance, in the typical
  point transistor, an emitter-to-base voltage may be served as such a parameter.
  Indeed, the electron flow across this device is an oscillating function of
  this voltage. In this regard, it is of interest to construct the point 
  interactions  with a resonance set controllable by parameters applied 
  externally and this is the main goal of the present paper. 
  
  In conclusion, in the present paper we have tried to develop the general approach
  on how to realize the point interactions as a zero-thickness limit of 
  structures composed of an arbitrary number of layers with biased potentials.
  This approach is specified by the examples describing one layer, 
  the double- and three-layer systems. The piecewise linear potentials 
  are not required to have any distributional limit as $\varepsilon \to 0$.
  Despite this, the $\varepsilon \to 0$ limit of the transmission matrices
  has been  shown to exist enable us to compute analytically the transmission 
  amplitude. The most interesting phenomenon discussed in the present paper
  is the appearance of the resonant transmission through a $\delta$-like
  barrier in the presence of an adjacent well. The origin of this effect
  emerges from the fact that the particle transmission across a well has an
  oscillating behavior. This behavior keeps to be of the same nature after tunneling 
  through a barrier. Therefore in the squeezed limit this oscillating 
  transforms into the function with non-zero values only at discrete points, whereas
  on the intervals between these points, this function converges pointwise to
  zero resulting in blocking the tunneling trough the barrier.

\bigskip 
{\bf  Acknowledgments}
\bigskip

One of us (A.V.Z) acknowledges partial financial support from 
the National Academy of Sciences of Ukraine (project No.~0117U000238).
G.P.T.  acknowledges support by the European Commission under project NHQWAVE 
(MSCA-RISE 691209). Y.Z. acknowledges  support from the 
Department of Physics and Astronomy
of the National Academy of Sciences of Ukraine under project No.~0117U000240.
%The authors are indebted to Fabio Rinaldi and Jose M. Mu\~{n}oz-Casta\~{n}eda
%for their suggestions and
%corrections resulting in the significant improvement of the work.

 \end{document}